\documentclass[aps,floats]{revtex4}
\usepackage{amsmath}
\usepackage{graphicx,epsfig}

\begin{document}
\bibliographystyle {plain}

\def\oppropto{\mathop{\propto}} 
\def\opsimeq{\mathop{\simeq}}
\def\opoverderline{\mathop{\overline}}
\def\operarrow{\mathop{\longrightarrow}}
\def\opsim{\mathop{\sim}}

\def\fig#1#2{\includegraphics[height=#1]{#2}}
\def\figx#1#2{\includegraphics[width=#1]{#2}}


\title{ Probing the tails of the ground state energy distribution \\
for the directed polymer in a random medium of dimension $d=1,2,3$  \\
via a Monte-Carlo procedure in the disorder   } 


\author{ C\'ecile Monthus and Thomas Garel }
 \affiliation{Service de Physique Th\'{e}orique, CEA/DSM/SPhT\\
Unit\'e de recherche associ\'ee au CNRS\\
91191 Gif-sur-Yvette cedex, France}

\begin{abstract}

In order to probe with high precision the tails of the 
ground-state energy distribution 
of disordered spin systems,
 K\"orner, Katzgraber and Hartmann \cite{Ko_Ka_Ha} have recently proposed
 an importance-sampling Monte-Carlo Markov chain in the disorder.
 In this paper, we combine their Monte-Carlo 
procedure in the disorder with exact transfer matrix calculations
in each sample to measure the negative tail of ground state
energy distribution $P_d(E_0)$
for the directed polymer in a random medium of dimension $d=1,2,3$.
In $d=1$, we check the validity of the algorithm by a direct comparison
with the exact result, namely the Tracy-Widom distribution.
In dimensions $d=2$ and $d=3$, we measure the negative tail up to
ten standard deviations, which correspond to 
probabilities of order $P_d(E_0) \sim 10^{-22}$.
Our results are in agreement with Zhang's argument,
stating that the negative tail exponent $\eta(d)$ of
the asymptotic behavior
 $\ln P_d (E_0) \sim - \vert E_0 \vert^{\eta(d)}$ as $E_0 \to -\infty$
is directly related to the fluctuation exponent $\theta(d)$
( which governs the fluctuations $\Delta E_0(L) \sim L^{\theta(d)}$
of the ground state energy $E_0$ for polymers of length $L$)
via the simple formula $\eta(d)=1/(1-\theta(d))$.
Along the paper, we comment on the similarities and differences with
spin-glasses.

\bigskip


\end{abstract}

\maketitle

\section{Introduction }

Since the ground-state energy $E_0$ of a disordered sample
is the minimal energy among the energies of all possible configurations,
the study of its distribution belongs to the field of extreme value statistics.
Whereas the case of independent random variables is well 
classified in three universality classes \cite{Gum_Gal}, the problem for 
the correlated energies within a disordered sample remains open
and has been the subject of many recent studies. 
The interest lies both \\
(i) in the scaling behavior of the average
 $E_0^{av}(L)$ and the standard deviation $ \Delta E_0(L)$ 
 with the size $L$ \\
(ii) in the asymptotic distribution $P(x)$ 
of the rescaled variable $x=(E_0 -E_0^{av}(L))/\Delta E_0(L)$
in the limit $L \to \infty$
\begin{equation}
{\cal P}_L(E_0)   \opsimeq_{L \to \infty}  \frac{1}{ \Delta E_0(L)} \  
P \left( x= \frac{
E_0 -E_0^{av}(L)}{ \Delta E_0(L) }  \right) 
\label{scalinge0}
\end{equation}
In this introduction, we first recall what is known
in the field of spin-glasses, before focusing
on the directed polymer model.

\subsection{ Ground state energy distribution in spin-glasses}

For spin-glasses in dimension $d$, let us consider samples 
containing $N=L^d$, where 
where $L$ denotes the linear size, and follow the
notations of Ref. \cite{Bou_Krz_Mar}.
The ` shift exponent' $\theta_s$ governs the correction to extensivity
of the averaged value
\begin{eqnarray}
E_0^{av}(L) \sim L^d e_0+ L^{\theta_s} e_1 +...
 = N e_0 + N^{\theta_s/d} e_1+...
\label{e0av}
\end{eqnarray}
Within the droplet theory \cite{Fis_Hus,Fis_Hus_SG},
this shift exponent $\theta_s$ coincides
with the domain wall exponent $\theta_{DW}$ and
with the droplet exponent $\theta$ of low energy excitations.
The ` fluctuation exponent' $\theta_f $  governs the growth
 of the standard deviation
 \begin{eqnarray}
\Delta E_0(L)  \sim  L^{\theta_f} e_2 = N^{\theta_f/d} e_2
\label{deltae0}
\end{eqnarray}
In any finite dimension $d$, it has been proven
that the fluctuation
exponent is $\theta_f=d/2$ \cite{We_Ai}.
Accordingly, the rescaled distribution $P(x)$ of Eq. (\ref{scalinge0}) 
was numerically found to be Gaussian in $d=2$ and $d=3$ 
\cite{Bou_Krz_Mar},
suggesting some Central Limit theorem.
On the contrary, in mean-field spin-glasses, the width
does not grows as $N^{1/2}$ and the distribution is not Gaussian.
In the Random Energy Model \cite{rem}, the width remains finite
$\Delta E_0(N) \sim O(1)$ and the distribution is the Gumbel distribution
\cite{bou_mez}. In the Sherrington-Kirpatrick model, the width grows
as $\Delta E_0(N) \sim N^{1/4}$ according to some theoretical
arguments \cite{asp_moo,Bou_Krz_Mar} and numerics 
\cite{ Bou_Krz_Mar,palassini}, and the distribution is clearly
asymmetric \cite{palassini,Ko_Ka_Ha}.
Finally for the one dimensional disordered spin chain with power-law
interactions that allows to interpolate between effectively
finite-dimensional and mean-field models,
the transition between short-range and infinite-range behaviors
corresponds to the Gaussian-non Gaussian transition for the
ground state energy \cite{germany}.

\subsection{ Ground state energy for the directed polymer}

The directed polymer model in $1+d$ dimensions,
 is defined by the following recursion for the partition function
\begin{eqnarray}
Z_{L+1} (\vec r) =  \sum_{j=1}^{2d}
 e^{-\beta \epsilon_L(\vec r+\vec e_j,\vec r)} Z_{L} (\vec r+\vec e_j)
\end{eqnarray}
The bond-energies $\epsilon_L(\vec r+\vec e_j,\vec r) $
are random independent variables, drawn with
 the Gaussian distribution
\begin{eqnarray}
\rho (\epsilon) = \frac{1}{\sqrt{2\pi} } e^{- \frac{\epsilon^2}{2} }
\end{eqnarray}
This model has attracted a lot of attention
for two main reasons : (i)  it is directly related
to non-equilibrium properties of growth models 
\cite{Hal_Zha} (ii) as a disordered system,
 it presents some similarities with the spin-glass physics
\cite{Hal_Zha,Der_Spo,Der,Mez,Fis_Hus}.  At low
temperature, there exists a disorder dominated phase, where the 
order parameter is an `overlap' \cite{Der_Spo,Mez}.

The probability distribution of the ground state energy $E_0$
is expected to follows the scaling form of Eq. (\ref{scalinge0}).
In contrast with spin-glasses where the shift exponent $\theta_s$
 (Eq. \ref{e0av})
and the fluctuation exponent $\theta_f$ 
 (Eq. \ref{deltae0}) are different,
there is a single exponent $\theta(d)$ that governs both
 the correction to extensivity
of the average $E_0^{av}(L)$ and the width $\Delta E_0(L)$ 
\begin{eqnarray}
E_0^{av}(L) && \sim L e_0 + L^{\theta(d)} e_1 +... \\
\Delta E_0(L) && \sim  L^{\theta(d)} e_2 +...
\end{eqnarray}
This exponent also governs the statistics of low excitations
within the droplet theory \cite{Fis_Hus}, as confirmed numerically
\cite{DPexcita}.
This exponent
is exactly known in one-dimension \cite{Hus_Hen_Fis,Kar,Joh,Pra_Spo}
\begin{eqnarray}
\theta(d=1)=1/3
\label{omegad1}
\end{eqnarray}
and has been numerically measured in dimensions
 $d=2,3,4,5$
\cite{Tan_For_Wol,Ala_etal,KimetAla,Mar_etal,DPexcita}
\begin{eqnarray}
\theta(d=2) && \sim 0.244 \\
\theta(d=3) && \sim 0.186
\label{omegad2et3}
\end{eqnarray}
For the mean-field version on the Cayley tree,
the exponent vanishes
 $\theta(d=\infty)=0$ \cite{Der_Spo,Dea_Maj},
with a width of order $O(1)$ for the probability distribution,
 but with a non random $O(\ln L)$
correction to the extensive term $e_0 L$ in the averaged value
  \cite{Der_Spo}.

The rescaled distribution $P_d$ is exactly known
in $d=1$ and is related to Tracy-Widom distributions
of the largest eigenvalue of random matrices ensembles
 \cite{Joh,Pra_Spo,prae}. On the Cayley tree,
the rescaled distribution was found to be non universal
and to depend on the disorder distribution \cite{Dea_Maj}.

\subsection{ Numerical measure of the ground state energy distribution}

The numerical measure of the ground state energy distribution 
is usually done by a simple sampling procedure, where 
the histogram of the energies of independent samples are collected.
However recently, K\"orner Katzgraber and Hartmann \cite{Ko_Ka_Ha} 
have proposed an importance-sampling 
Monte-Carlo algorithm in the disorder, which allows to measure
much more precisely the tails of the distribution.
In the case of the Sherrington-Kirkpatrick model of spin-glasses,
this procedure was used to measure the negative tail 
on systems of size $N \leq 128$ \cite{Ko_Ka_Ha}
up to $ x \geq - 15 $ corresponding to probabilities $P(x) \geq 10^{-18}$
(see Eq. \ref{scalinge0}), whereas the simple sampling procedure
cannot go beyond $ x \geq -5$ corresponding to  $P(x) \geq 10^{- 4}$ \cite{palassini}.

For the directed polymer 
in dimensions $d=2$ and $d=3$, the rescaled distribution $P_d$
has been numerically measured via simple sampling in \cite{Kim_Moo_Bra}
with results in the region $ x \geq -5 $.
In this paper, we use the importance sampling algorithm 
recently proposed in \cite{Ko_Ka_Ha} to measure precisely
the negative tail of the probability distribution
up to $x \geq -10$.

The paper is organized as follows.
In Section \ref{zhangargument}, we recall Zhang's argument \cite{Hal_Zha}
that relates the decay of the rescaled distribution $P_d$
to the fluctuation exponent $\theta_f$.
In Section \ref{montecarlo}, we describe the Monte-Carlo procedure
in the disorder proposed in \cite{Ko_Ka_Ha} and mention
the specific choices for the application to the directed polymer model.
In Section \ref{res1d}, we show the validity of the procedure
in $d=1$ via the direct comparison with the exactly known distribution
(Tracy-Widom). Finally in Sections \ref{res2d} and \ref{res3d},
we present our results for $d=2$ and $d=3$ respectively.
We present our conclusions in Section \ref{conclusion}.

\section{Zhang's argument for the negative tail exponent }

\label{zhangargument}

\subsection{ Distribution of the free-energy in the low temperature phase }

According to the droplet theory\cite{Fis_Hus}, 
the whole low temperature phase $0<T<T_c$
is governed by a zero-temperature fixed point.
In particular, at $T<T_c$, the droplet exponent $\theta(d)$ governs 
the width $\Delta F(L,T)$ and the correction to extensivity
of the average $F^{av}(L,T)$
\begin{eqnarray}
\Delta F(L,T) && \sim  L^{\theta(d)} f_2(T) +... \\
F^{av}(L,T) && \sim L f_0(T) + L^{\theta(d)} f_1(T) +...
\end{eqnarray}
and the rescaled probability distribution of the free-energy
 coincides with the rescaled distribution
$P_d$ describing the ground-state energy distribution (\ref{scalinge0})
\begin{equation}
{\cal P}_d(F,L,T)   \simeq  \frac{1}{ \Delta F(L,T)  }   \ 
P_d \left( x= \frac{
F - F^{av}(L,T)}{ \Delta F(L,T) }  \right) 
\label{rescalingfreelow}
\end{equation}
as recently checked numerically using simple sampling \cite{DPtransi}.

\subsection{Zhang's argument for the directed polymer }

In finite dimensions $d>1$, the rescaled distribution $P_d$ is not known
 but there exists a simple argument due to Zhang \cite{Hal_Zha}
 that allows to determine
the exponent $\eta$ of the negative tail of the free energy distribution
\begin{eqnarray}
P_d( x \to -\infty) \sim e^{- c \vert x \vert^{\eta(d)} } 
\label{taileta}
\end{eqnarray}
If $\eta(d)>0$, the moments of the partition function can be 
 evaluated by the saddle-point
method, with a saddle value $F^*$ lying in the negative tail (\ref{taileta})
\begin{eqnarray}
\overline{ Z_L^n} = \int dF {\cal P}_L(F,L) e^{ - n \beta F}  \sim \int dF 
e^{- c \left( \frac{ \vert F \vert}{L^{\theta(d)}} \right)^{\eta(d)} }
 e^{ - n \beta F} 
\sim e^{ b(n) L^{ \frac{ \theta(d) \eta(d)}{\eta(d)-1}   } }
\label{saddle}
\end{eqnarray}
Since for positive integer $n$, these moments of the partition function 
can be formulated in terms of the iteration of some transfer matrix,
they have to diverge
 exponentially in $L$ with some Lyapunov exponent.
As a consequence, the exponent $\eta(d)$ of the negative tail (\ref{taileta})
 is not a free parameter, 
but is fixed by the value of the fluctuation exponent 
\begin{eqnarray}
\eta(d)=\frac{1}{1-\theta(d)} 
\label{etad}
\end{eqnarray}
In dimension $d=1$ where the droplet exponent 
is exactly known $\theta(d=1)=1/3$ (Eq. \ref{omegad1}),
this yields the negative tail exponent
\begin{eqnarray}
\eta(d=1)= \frac{3}{2} 
\label{etad1}
\end{eqnarray}
in agreement with the exact Tracy-Widom distributions \cite{Joh,Pra_Spo,prae}.
In dimensions $d=2$ and $d=3$, the numerical estimates 
of the droplet exponents (Eq. \ref{omegad2et3}) yield the following predictions
\begin{eqnarray}
\eta(d=2) && \sim 1.32 \nonumber \\
\eta(d=3) && \sim 1.23
\label{etad2et3}
\end{eqnarray}
These predictions have been tested numerically in \cite{Kim_Moo_Bra}
using simple sampling that do not allow to have data far in the tails.
In the following, we will use the importance sampling Monte-Carlo
method in the disorder to probe the negative tail more precisely.

\subsection{Zhang's argument for spin-glasses }

To the best of our knowledge, Zhang's argument seems to be applied
only in the context of directed polymers \cite{Hal_Zha},
whereas it can be applied for other kinds of disordered systems
since it is only based on scaling argument within
a saddle-point approximation in the large $L$ limit (Eq. \ref{saddle}).
It is thus interesting to describe now its implications
in the field of spin-glasses.

For spin-models in finite dimensions, the fluctuations
of free energies over the samples scale instead as
\begin{eqnarray}
\left[ \Delta F_L \right]_{samples}  \sim  L^{d/2}
\label{clt}
\end{eqnarray}
at any temperature as proven in \cite{We_Ai}.
This scaling simply reflects the Central-Limit fluctuations 
of the $L^d$ disorder variables
defining the sample. ( The directed polymer escapes
 from these normal fluctuations
because it is a one-dimensional path living 
in a $1+d$ disordered sample :
each configuration of the polymer only sees $L$ random variables
among the $L^{1+d}$ disorder variables that define the sample,
and the polymer can 'choose' the random variables it sees.)

Repeating Zhang argument in this case (\ref{clt}) yields for the negative
 tail exponent $\eta(d)=2$.
This is in agreement with the recent numerical studies
\cite{Bou_Krz_Mar,germany} that find a Gaussian distribution
in finite dimensions and in the one dimensional Ising spin-glass with
long range interactions in the non-mean-field regime.

On the contrary, for the Sherrington-Kirkpatrick model
\cite{palassini,Bou_Krz_Mar,Ko_Ka_Ha},
the probability distribution of the ground state
is found to be asymmetric, and has been fitted
with generalized Gumbel distribution \cite{palassini,Ko_Ka_Ha}.

However, if one repeats Zhang argument for the SK model
with the measured fluctuation exponent $\theta_f \sim 0.235$
\cite{palassini} for the width $\Delta E_0(N) \sim N^{\theta_f}$,
one obtains the negative tail exponent 
\begin{eqnarray}
\eta_{SK} = \frac{1}{1-\theta_f} \sim 1.3
\label{etaSK}
\end{eqnarray}
If the value of the fluctuation exponent is exactly $\theta_f =1/4$
as suggested by some theoretical
arguments \cite{asp_moo,Bou_Krz_Mar}, the negative tail exponent
would be $\eta_{SK}=4/3$.

This could explain why the fit with generalized Gumbel distributions
whose negative tail 
is a simple exponential $e^{-m \vert x \vert}$
  with exponent $\eta=1$ and coefficient $m$
leads to increasing effective values of $m$ when the 
range over which the tail is measured grows :
the fit with simple scaling data on $ x \geq -6$
leads to
$m \sim 6$ \cite{palassini},
whereas the importance scaling data on $ x \geq -15$
leads to a completely different estimate $m \sim 11 $ \cite{Ko_Ka_Ha}.

\section{Description of the importance-sampling Monte-Carlo
algorithm in the disorder }

\label{montecarlo}

In Ref. \cite{Ko_Ka_Ha}, a procedure based on an 
importance-sampling Monte-Carlo
algorithm in the disorder was proposed to probe with high precision
the tails of the ground-state energy distribution of disordered systems,
and was applied for the Sherrington-Kirpatrick mean-field Ising
 spin-glass, where probabilities up to $10^{-18}$ could be measured.
In this Section, we summarize their method which can be divided in
three steps. For each step we mention the specific choices
we have made to apply it to the directed polymer model.

\subsection{ Simple sampling }

\begin{figure}[htbp]
\includegraphics[height=4cm]{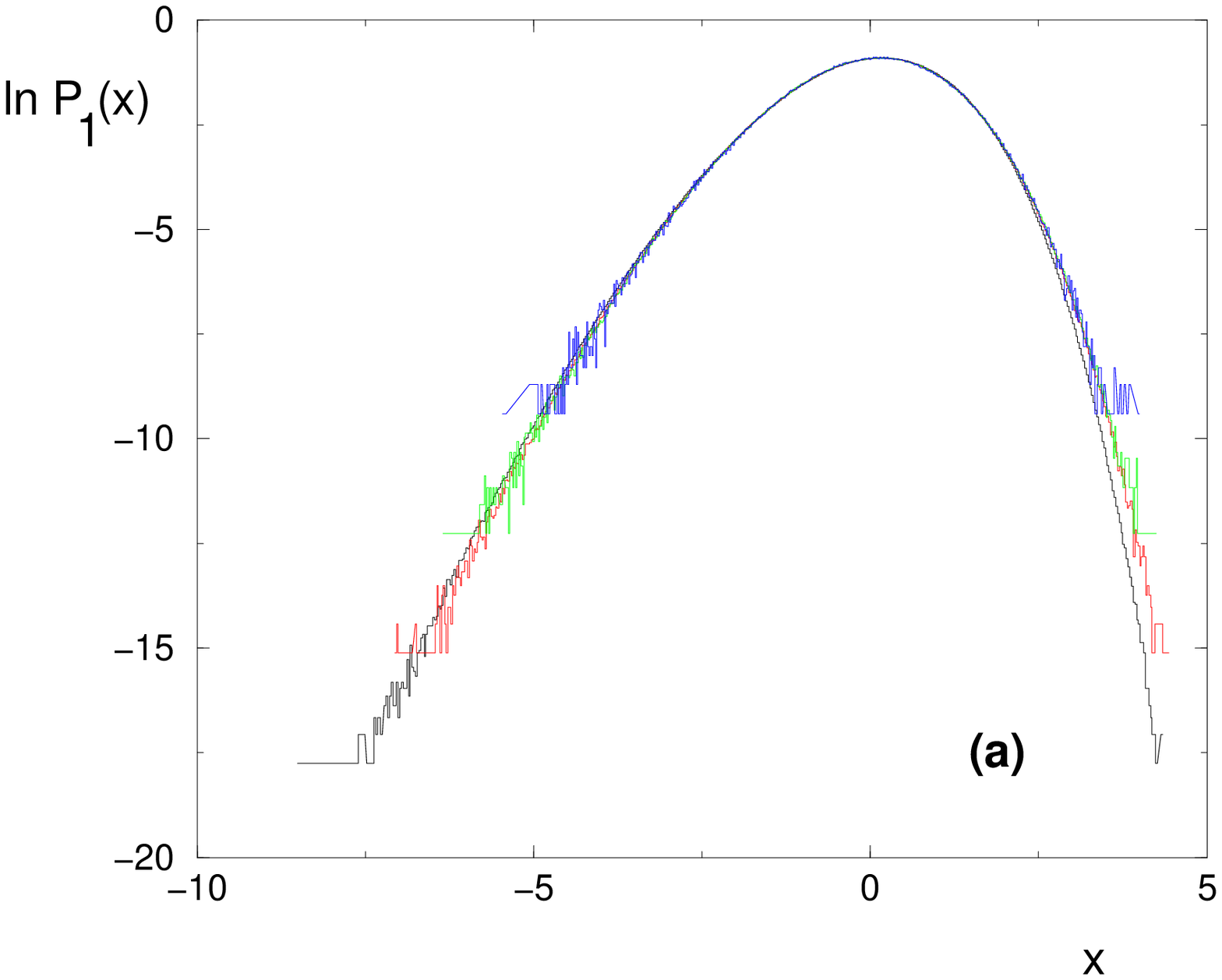}
\hspace{0.5cm}
\includegraphics[height=4cm]{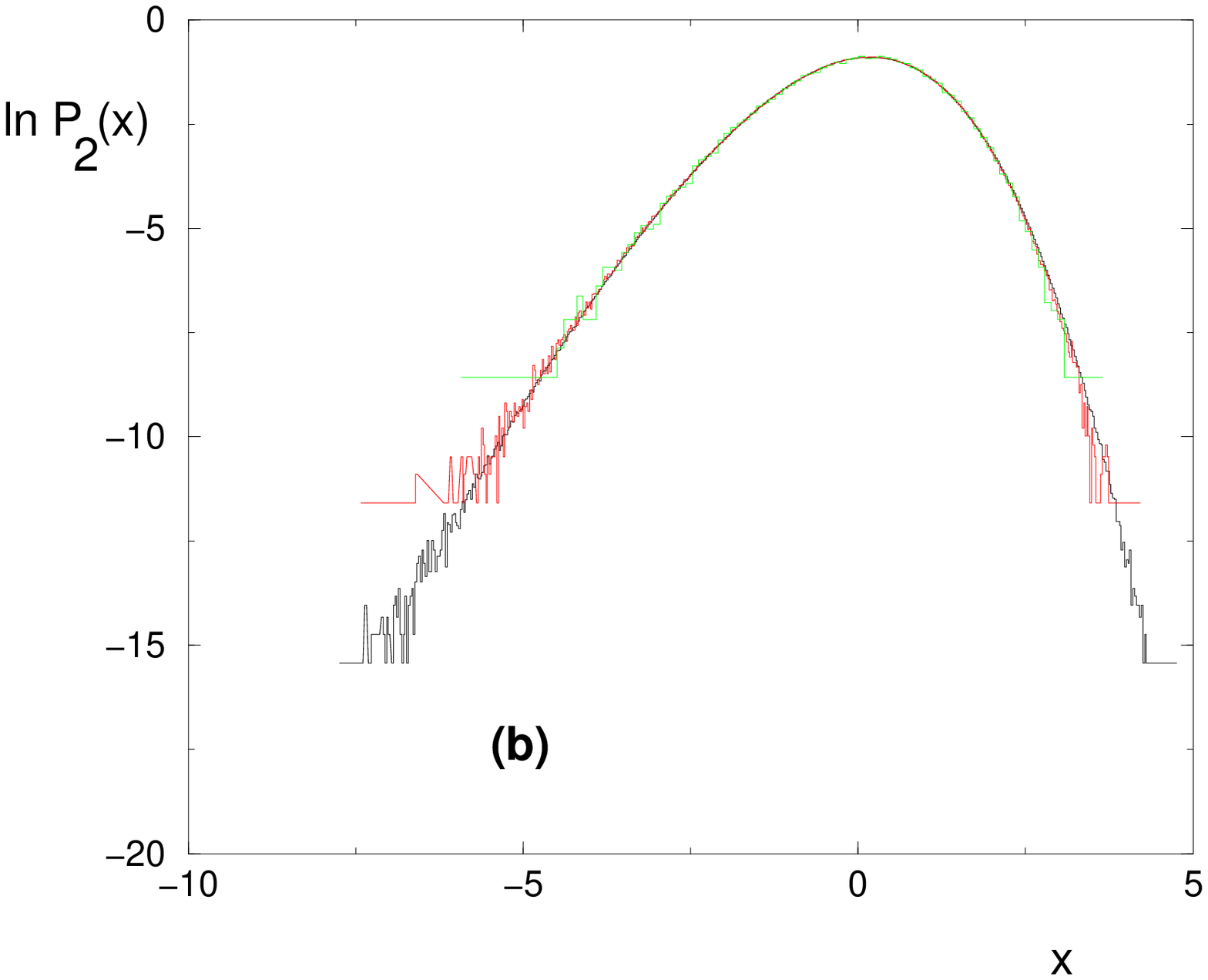}
\hspace{0.5cm}
\includegraphics[height=4cm]{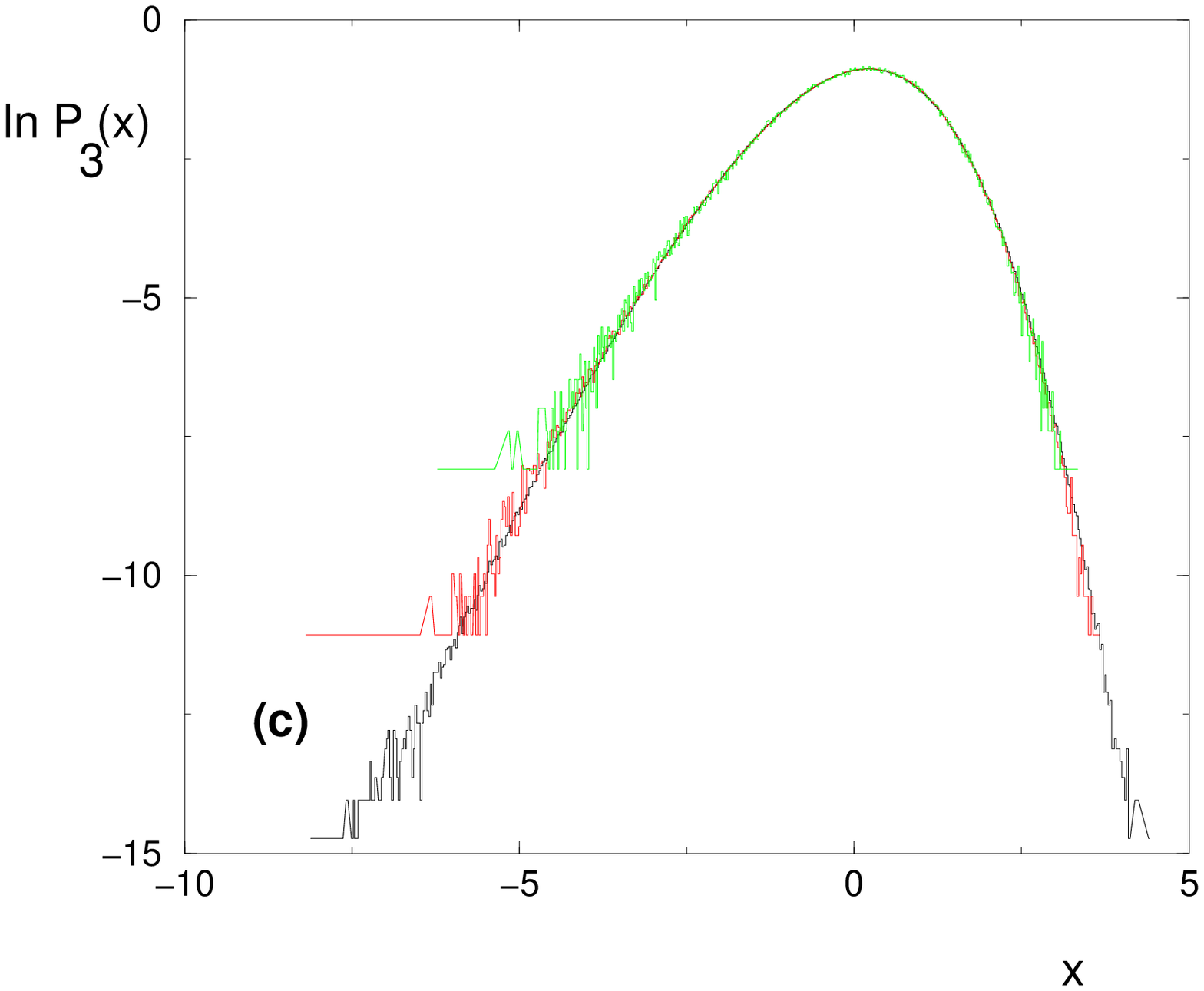}
\caption{ (Color online)  Logarithmic plot of the rescaled
 probability distribution 
$P_{d}(x)$ (Eq. (\ref{scalinge0})) as measured via simple sampling for
(a) $L=12,50,200,800$ in $d=1$ ;
(b) $L=10,40,160$ in $d=2$;
(c) $L=6,18,36$ in $d=3$ }
\label{figsimple}
\end{figure}

A disorder configuration will be denoted by $\cal D$, and its ground state
energy $E(\cal D)$. 
For the directed polymer, the ground state energy can be computed via transfer
matrix.
A simple sampling numerical estimation $P_{simple}(E)$ 
of the ground state energy distribution $P(E)$
consists in drawing $n_s$ independent disordered samples ${\cal D}_1,...
{\cal D}_{n_s}$, in computing the corresponding ground state energies
$E({\cal D}_1),...,E({\cal D}_{n_s})$, and in constructing the histogram

\begin{eqnarray}
\label{simple}
P_{simple}(E) = \frac{1}{n_s} 
\sum_{i=1}^{n_s} \delta \left(E-E({\cal D}_i) \right)
\end{eqnarray}
This histogram is very useful to measure the distribution $P(E)$
where $P(E) \gg \frac{1}{n_s}$, but gives no information
on the tails where $P (E) < \frac{1}{n_s}$, since no events are found.

As an example, we show on Fig. \ref{figsimple}
the results we have obtained recently
via simple sampling for $d=1,2,3$ respectively \cite{DPexcita} :
whereas the core of the distribution is well measured,
the tails suffer from statistic fluctuations as soon
as the probability becomes too small.
Moreover, if one chooses to make the same CPU effort on all sizes,
the number of samples rapidly decay with the size $L$,
so that the data for the tails are less and less precise as $L$ grows.
Since one is interested
into the asymptotic regime $L \to \infty$, the correct measure
of the tails quickly becomes intractable within the simple sampling procedure.

This is why a correct measure of the tails requires the use of
some importance sampling, as stressed in \cite{Ko_Ka_Ha}.
 However, the simple sampling study
is the first necessary step within the present method,
for three reasons : \\
(i) the simple sampling results are needed to construct
the guiding function of the importance sampling measure\cite{Ko_Ka_Ha} 
as described below  \\
(ii) the simple sampling results give accurate results for the
average value $E_0^{av}(L) $ and the standard deviation $ \Delta E_0(L) $,
that do not have to be measured via importance sampling \cite{Ko_Ka_Ha}.
In particular, this allows to work on a finite box $[x_{min},x_{max}]$
for the rescaled variable (\ref{scalinge0}), and to choose freely
the boundaries of the box, for instance $x_{max}=-1.$ to concentrate
on the negative tail, as will be done below for the directed polymer.   \\
(iii) finally, the simple sampling results allow to check
the validity of the importance sampling measures on the core of
the distribution where the simple sampling results are sufficiently precise.

\subsection{ Construction of a guiding function $G(E)$
 from the simple sampling result $P_{simple}(E)$}

\begin{figure}[htbp]
\includegraphics[height=4cm]{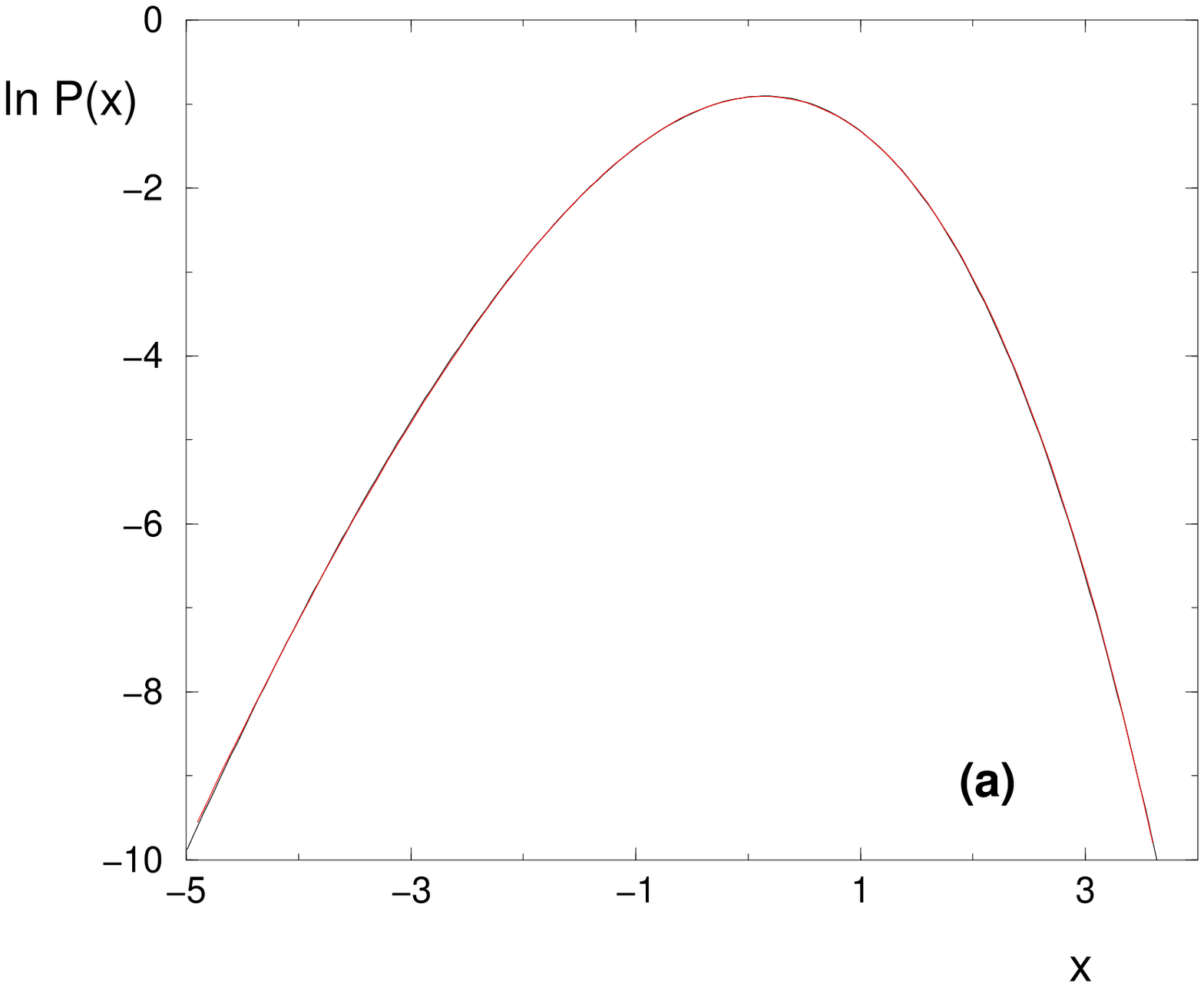}
\hspace{0.5cm}
\includegraphics[height=4cm]{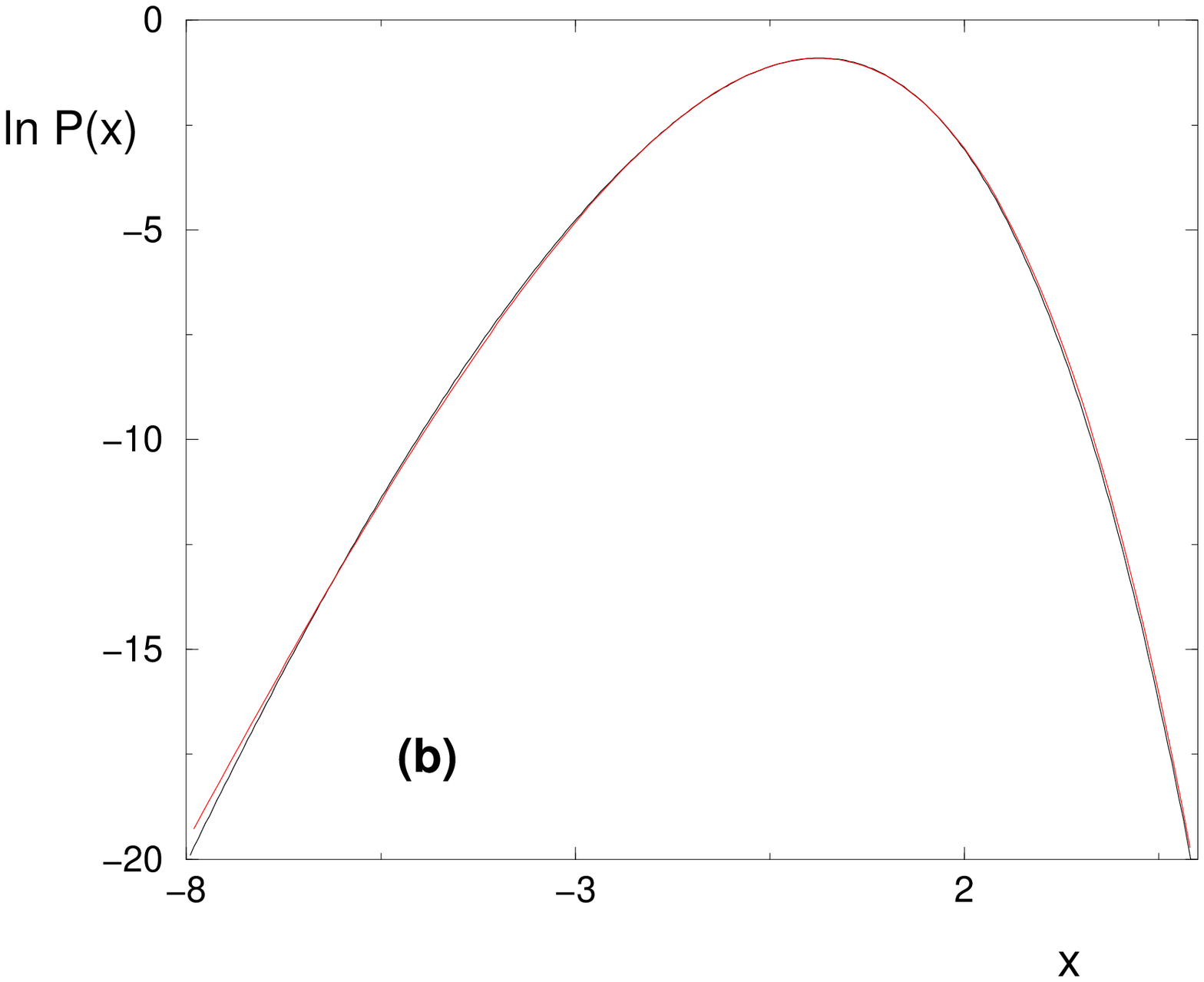}
\hspace{0.5cm}
\includegraphics[height=4cm]{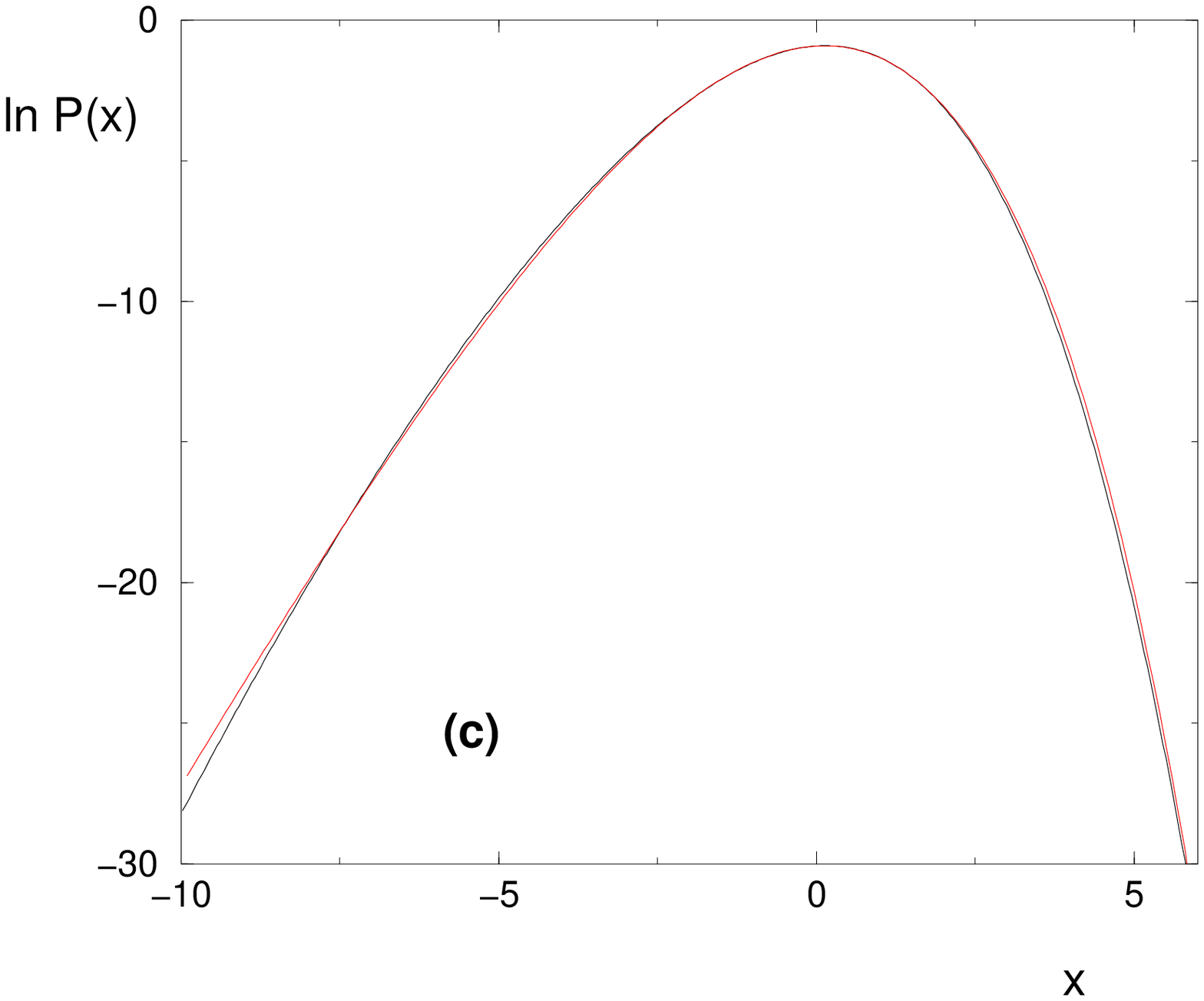}
\caption{ (Color online) Fits of the GOE
 Tracy-Widom distribution $ \ln P_{TW}^{GOE}(x)$
by generalized
Gumbel distributions $\ln g_m(x)$ (Eq. \ref{genegumbel})
 on various intervals :
(a) fit on [-5,3.7] corresponding to $\ln P > -10$ with $m=12.93$ ;
(b) fit on [-8,5] corresponding to $\ln P > -20$ with $m=14.71$
(c) fit on [-10,6] corresponding to $\ln P > -30$ with $m=15.92$ ;
 }
\label{figtracy}
\end{figure}

The simple sampling result $P_{simple}(E)$
exists in the range in $E$ where $P_{simple}(E) > 1/n_s$,
whereas the guiding function $G(E)$ needed for the importance
sampling below has to be defined in the tails where $P_{simple}(E) < 1/n_s$.
The guiding function $G(E)$ should be in some sense
the `best' extrapolation of the data $P_{simple}(E)$.
The proposal of \cite{Ko_Ka_Ha} is to define $G(E)$
as the best fit of $P_{simple}(E)$ within the one parameter family
of generalized Gumbel distribution $g_m(x)$, which reads
for the  normalization conditions $<x>=0$ and
$<x^2>=1$
\begin{equation}
\label{genegumbel}
g_m(x) \equiv \frac{1}{\beta(m)}\frac{m^m}{\Gamma(m)}
\left(e^{\frac{x-\alpha(m)}{\beta(m)}
-e^{\frac{x-\alpha(m)}{\beta(m)}}}\right)^m
 \end{equation}
where
$\beta(m)=\frac{1}{\sqrt{\frac{\Gamma''(m)}{\Gamma(m)}-\bigl(\frac{\Gamma'(m)}{\Gamma(m)}\bigr)^2}}$ 
and $\alpha(m)=-\beta(m) \left(\frac{\Gamma'(m)}{\Gamma(m)}-\rm {ln} \
m \right)$. The usual Gumbel distribution corresponds to $m=1$,
whereas the Gaussian can be formally recovered in the limit $m \to \infty$.
This choice was motivated by the numerical finding that the rescaled
probability distribution as measured via simple sampling
could be fitted extremely well by a generalized Gumbel distribution $g_m(x)$
of parameter $m \sim 6$ \cite{palassini,Ko_Ka_Ha}.
It turns out that many recent studies in various contexts
 have found that asymmetric distributions could be extremely
well fitted by generalized Gumbel distributions with various non-integer 
values of $m$
\cite{Bramwell,BBW,Rosso, interface}. This has motivated theoretical
studies to understand the origin of this type of distribution \cite{Bertin}.
However as discussed in ref. \cite{Bramwell}, 
it is empirically known that probability distribution functions (PDF)
with the same first four moments approximately coincide over the range
of a few standard deviation which is precisely the range of numerical
-or experimental- data. So generalized Gumbel PDF's, with arbitrary $m$, 
should not be considered more than a convenient one
parameter fit. To demonstrate clearly how misleading
these fits can be, we show on Fig \ref{figtracy}
how the Tracy-Widom GOE distribution $P_{TW}^{GOE}$
(which represents the exact rescaled
distribution for the directed polymer model in $d=1$ \cite{Joh,Pra_Spo,prae})
can be fitted by generalized Gumbel distributions 
on the three ranges $P> 10^{-10}$, $P> 10^{-20}$ and $P> 10^{-30}$ :
the best fit corresponds to increasing values of the parameter $m$.
For the first range $P> 10^{-10}$, the found fit is `perfect',
whereas a slight difference begin to appear in the negative tail
as the range grows. Moreover, the Tracy-Widom distribution
is known to have the following asymptotic behavior 
\begin{eqnarray}
\label{tracyasymp}
P_{TW}^{GOE}(x) \opsimeq_{x \to -\infty} e^{- c_1 \vert x \vert^{\eta_1} }
\ \ { \rm  with } \ \ \eta_1=\frac{3}{2}
 \end{eqnarray}
whereas the generalized Gumbel distribution have for for any $m$
an exponential tail with exponent $\eta=1$ and coefficient $m$
\begin{eqnarray}
\label{gumbelasymp}
g_m(x) \opsimeq_{x \to -\infty}
e^{ -m \vert x \vert }
 \end{eqnarray}
This explains why the effective $m$ of the best fit grows with the range.
In conclusion, whenever the fit of the core of the distribution
leads to an effective $m$ which grows with
the range, as in the SK model where $m \sim 6$ and $m \sim 11$
were found depending on the range \cite{palassini,Ko_Ka_Ha},
 the PDF is probably not a generalized Gumbel distribution,
but is likely to have a negative tail exponent $\eta>1$
(as already suggested around Eq. \ref{etaSK} using Zhang's argument ).
And if one focuses on the negative tail, 
it is clear that the fit with a simple exponential (\ref{gumbelasymp})
is very restrictive.

As a consequence, in the following where we focus on the negative tail
$ x \leq -1$ for the directed polymer, we have chosen not to work
with generalized Gumbel distribution, but to construct a guiding
function $G(E)$ which fit the simple sampling data and whose leading behavior
involves the negative tail exponent $\eta_d$ as obtained
from Zhang argument ( see Eqs \ref{etad}, \ref{etad1} and \ref{etad2et3}).
In practice, we have found convenient to work in $d=2$ and $d=3$
on the range $x \in [-10,-1]$ with some guiding function
$G_d(x)$ of the form
\begin{eqnarray}
\label{guidingchoice}
\ln G_d(x) = a_0 - a_1 \vert x \vert^{\eta_d} + a_2 \ln \vert x \vert
 \end{eqnarray}
where the three parameters $a_i(d)$ were chosen to fit best the simple
sampling data.

\subsection{ Importance sampling with the guiding function $G(E)$ }

The importance sampling Monte Carlo algorithm proposed in \cite{Ko_Ka_Ha}
is defined by the following Markov chain :

(1) From the current disorder configuration ${\cal D}_i$,
construct a candidate ${\cal D}'$ for the next disorder configuration 
${\cal D}_{i+1}$
by replacing a subset of ${\cal D}_i$ chosen at random with new values
drawn with the original disorder distribution.
For a spin model of $N$ spins, this subset can be for instance
a single bond chosen at random, or all bonds connected
to a site chosen at random, so that the proposed change
in the ground state energy is of order $O(1)$ with
respect to a value of order $E_0(N) = N e_0+...$,
i.e. its relative order of magnitude is of order $1/N$ \cite{Ko_Ka_Ha}.
For the directed polymer studied here, we have chosen for this subset
the energies of a whole time-slice, i.e. all the disorder variables
seen by a given monomer. Then the proposed change
in the ground state energy is of order $O(1)$ with respect to a 
value of order $E_0(N) = N e_0+...$ as in spin models.

(2) Calculate the new ground state energy $E({\cal D}')$
and compare it with the previous ground state energy $E({\cal D}_i)$ 
using the guiding function $G(E)$ :
set ${\cal D}_{i+1}={\cal D}'$ with probability
\begin{eqnarray}
\label{accept}
p_{accept}\left({\cal D}' \vert {\cal D}_i \right)  =
 min \left[ \frac{ G(E({\cal D}_i)) }{G(E({\cal D}'))},1 \right]
\end{eqnarray}
and set ${\cal D}_{i+1}={\cal D}_i$ otherwise.

This Markov chain is expected to converge towards a stationary
state where a disorder configuration
${\cal D}$ is visited with probability $\propto 1/G(E({\cal D}))$. 
The stationary probability to visit a disorder configuration with energy $E$
is now given by the ratio 
\begin{eqnarray}
\label{stationary}
R_{stationary}(E)= \frac{P(E)}{G(E)}
\end{eqnarray}
If the guiding function $G(E)$ were the exact $P(E)$, 
this would corresponds to a flat-histogram sampling of $P(E)$. 
If $G(E)$ is just a reasonable extrapolation of the simple sampling result 
$P_{simple}(E)$, one expect to measure nevertheless much better
the tails of $P(E)$.

(3) Measurements from the Monte-Carlo procedure :
since successive configurations visited by 
a Monte-Carlo algorithm are not independent, one usually keeps only 
decorrelated configurations for
 the numerical measure $R_{importance}(E)$
 of the theoretical stationary solution $ R_{stationary}(E)$.
This means in practice that one should first estimate
some typical correlation time $\tau$ and use only every $\tau$th 
configuration 
\begin{eqnarray}
\label{importance}
R_{importance}^{(\tau)}(E)= \frac{1}{m_I} \sum_{j=1}^{m_I} 
 \delta \left(E-E({\cal D}_{i+ j \tau} ) \right)
\end{eqnarray}
where the number $m_I$ of measured is simply the ratio
$m_I=\frac{T}{\tau}$ of the total number $T$ of Monte-Carlo
iterations by the correlation time $\tau$.
For instance in \cite{Ko_Ka_Ha}, 
the time $\tau$ was chosen to be $\tau=4 \tau_e$
where $\tau_e$ is the time where the autocorrelation of
 the ground state energy
\begin{eqnarray}
\label{corre}
C(t)= \frac{< E_i E_{i+t} > - < E_i > < E_{i+t} >}{ <E_i^2>-<E_i>^2}
\end{eqnarray}
decays to $1/e$. For the SK model with $16 \leq N \leq 128$ spins,
the autocorrelation time was found to be of order of 400-700 MC steps
 \cite{Ko_Ka_Ha}.

For the directed polymer, we actually find that the histograms 
$R_{importance}^{(\tau)}(E)$ obtained for $\tau=1$ and $\tau \gg \tau_e$
coincide, except that the histograms with large $\tau$ contain more
noise since they are built out of less events.
From a theoretical point of view, one can justify this finding
as follows : if the total Monte-Carlo time $T$ is much bigger 
than the typical time $t_{cross}$
to cross the interval $[E_{min},E_{max}]$, then the average with
respect to the stationary measure should be equivalent to the time
average of the Monte-Carlo procedure where all times are kept 
\begin{eqnarray}
\int
dE f(E) P_{stationary}(E) = \frac{1}{T} \sum_{t=1}^T f(E(t))
\ \ {\rm for } \ \   T \gg t_{cross}
\label{ergodicity}
\end{eqnarray}
Indeed for a free random walk in a finite box,
it seems clear that one obtains the flat histogram via measuring
the positions at all times, instead of throwing away 
most of the times to have independence between two consecutive measures.
The quality of the convergence towards the stationary distribution
then depends on the number
\begin{eqnarray}
n_{cross} \sim \frac{T}{t_{cross}}
\label{ncross}
\end{eqnarray}
of crossings of the interval $[E_{min},E_{max}]$ during the total number $T$
of the Monte-Carlo, which should be large enough $ n_{cross} \gg 1$.

\subsection{ Summary of the procedure used for the directed polymer  }

In the following sections, we will present the results 
for the ground state energy distribution
obtained by combining \\
(i) the Monte-Carlo procedure in the disorder discussed above \\
(ii) the transfer matrix calculation of 
the ground state energy in each sample with a free boundary condition
for the end polymer.

We have chosen to focus on the negative tail,
by working on the same finite box $x \in [x_{min},x_{max}]$
in terms of the rescaled variable $x$ (Eq. \ref{scalinge0}) for all sizes $L$.
We now present our results for $d=1,2,3$ respectively.

\section{ Measure of the ground state energy distribution in $d=1$}

\label{res1d}

In $d=1$, the exact rescaled distribution of the ground state energy
is exactly known and corresponds to
the Tracy-Widom GOE distribution $P_{TW}^{GOE}$ \cite{Joh,Pra_Spo,prae}
if the last monomer is free, the case we consider here. 
(It would be the Tracy-Widom GUE distribution $P_{TW}^{GUE}$ 
if the last monomer were fixed at the origin).
We use this exact result to check the validity of the Monte-Carlo
procedure in the disorder and to describe its main properties.

\subsection{ Numerical details }

In dimension $d=1$, we have chosen to work on the interval
$ [x_{min},x_{max}]=[-11.,-1.]$ for the rescaled variable $x$ 
(Eq. \ref{scalinge0}), i.e. to probe the negative tail up to probabilities
of order $P_1(x) > 10^{-32}$. 
We now give the sizes $L$
we have studied, together with the standart deviation $\Delta E_0(L)$
measured by simple sampling and used in the rescaling of Eq. 
\ref{scalinge0} (the averaged values $E_0^{av}(L)$ can be found in 
our previous work \cite{DPexcita}),
 the corresponding 
number $T_L$ of Monte-Carlo iterations,
the acceptation rate 
$\tau_{acc}(L)$ of Monte-Carlo moves,
and the number $n_{cross}(L)$ of crossings of the box $ [x_{min},x_{max}]=
[-11.,-1.]$.
\begin{eqnarray}
L && =50, 100,200,400,800,1600 \\
\Delta E_0(L) && \sim 3.12,3.98, 5.04, 6.36 , 8.04 , 10.11          \\
T_L && = 33. 10^8 , 85.10^7, 225.10^6, 57.10^6 , 7.10^6 ,  4. 10^6 \\
\tau_{acc}(L) && \sim 0.54,  0.64, 0.72, 0.76, 0.82 , 0.86 \\
n_{cross} (L) && \sim  98.10^4, 224.10^3, 45.10^3, 8.10^3, 670, 230
 \end{eqnarray}

As $L$ grows, the proposed Monte-Carlo
moves $\Delta x$ in the rescaled variable $x$ (Eq. \ref{scalinge0})
 are smaller : this is why both the acceptation rate $\tau_{acc}(L)$ 
 and the crossing time $t_{cross}(L)$ (Eq. \ref{ncross}) also
grows with $L$. The final result is that the number of crossing
$n_{cross}(L)$ decays with $L$, and since it should remain
large enough to obtain a good measure (Eq. \ref{ncross}),
this number fixes the maximal size that can be correctly
studied.

\subsection{ Properties of the Monte-Carlo process in the disorder }

\begin{figure}[htbp]
\includegraphics[height=6cm]{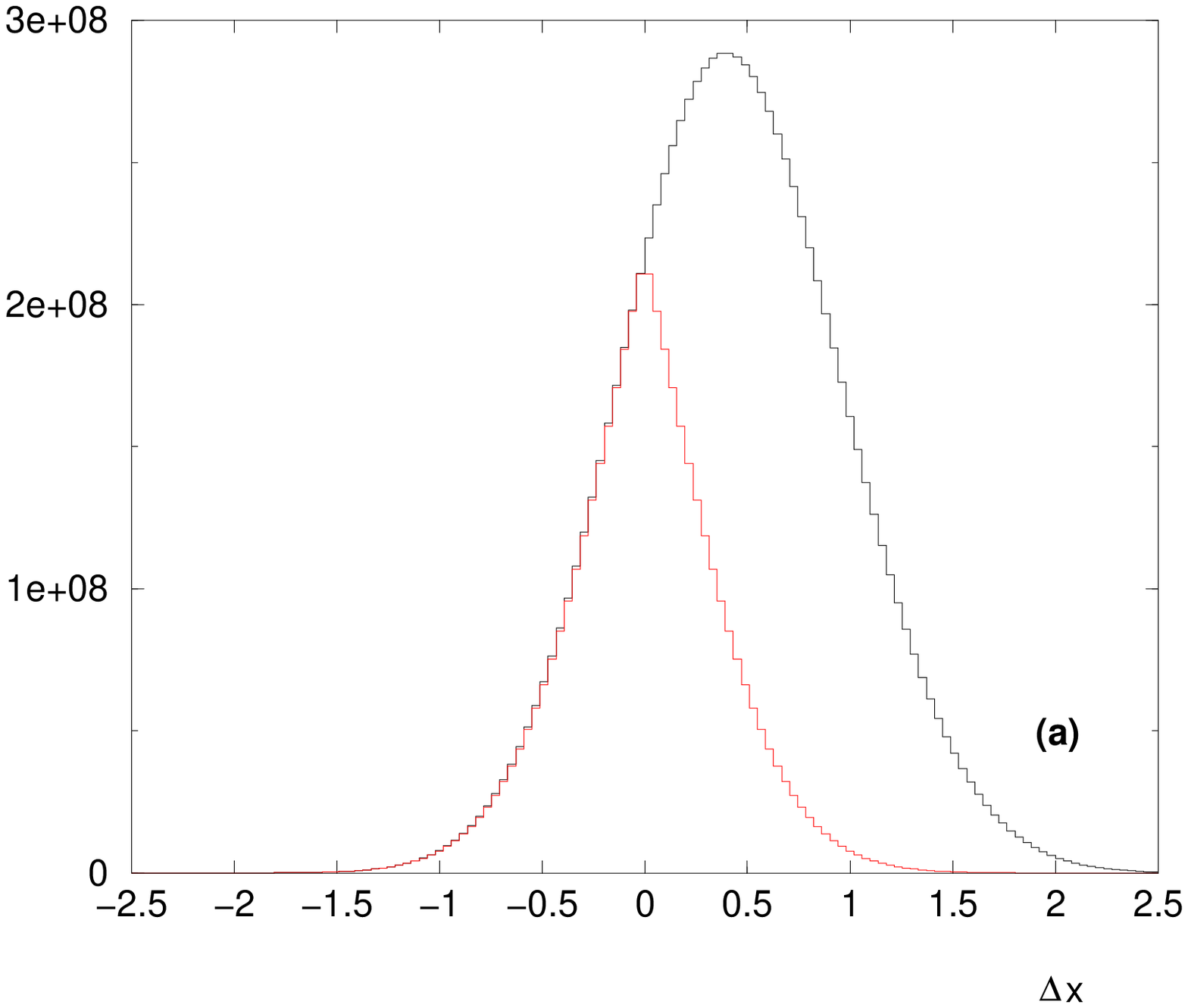}
\hspace{1cm}
\includegraphics[height=6cm]{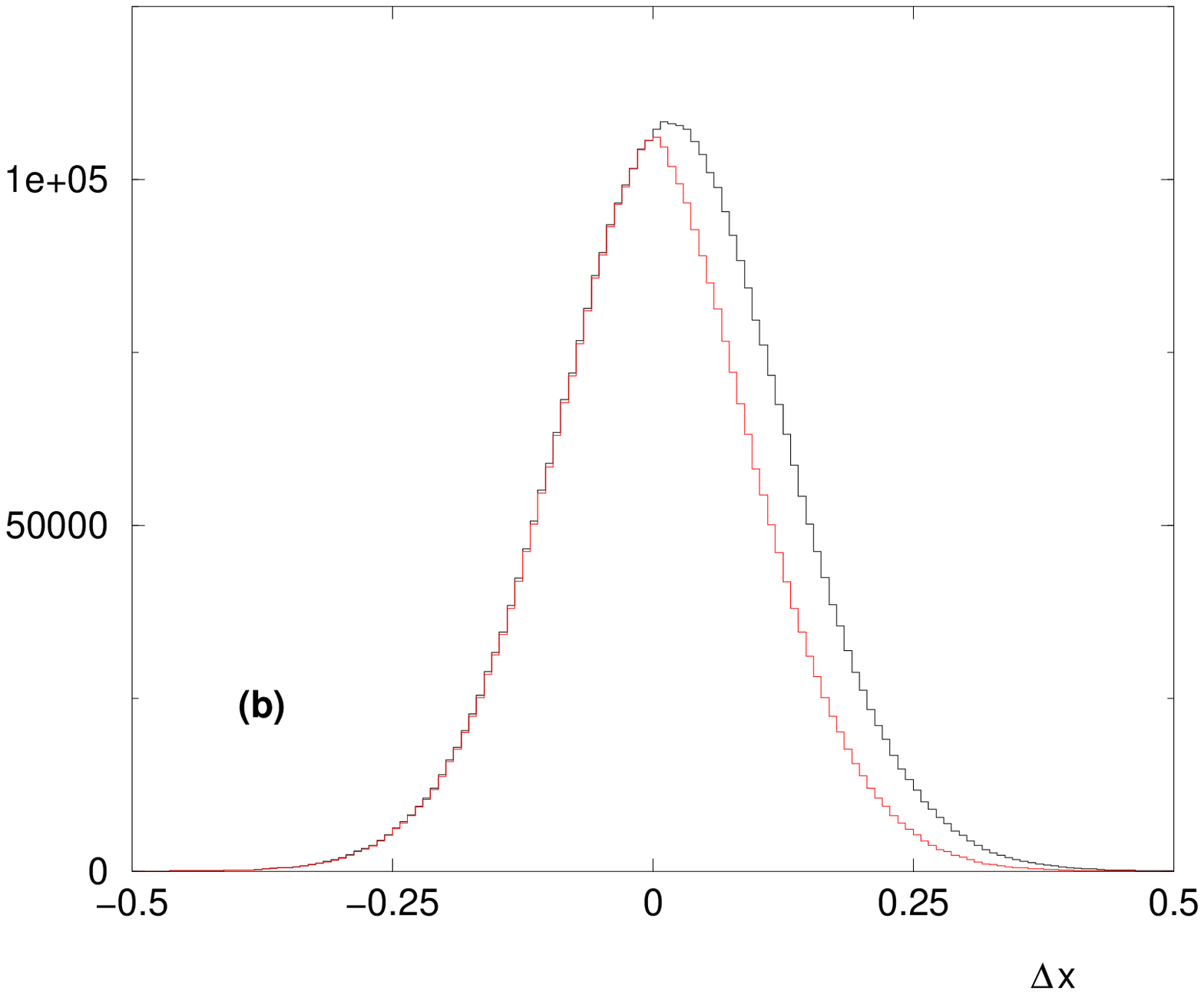}
\caption{ (Color online)  Monte-Carlo procedure to measure
 the negative tail on $x \in [-11,-1]$ in $d=1$  :
histograms of the proposed and accepted Monte-Carlo changes $\Delta x$
in the disorder
(a) for $L=25$ where the acceptation rate is $\tau_{acc} \sim 0.425$
(b) for $L=1600$ where the acceptation rate is $\tau_{acc} \sim 0.86$ . }
\label{fig1d5253}
\end{figure}

We show on Fig. \ref{fig1d5253} the histograms of the 
 proposed and accepted Monte-Carlo changes
in the disorder, for $L=25$ and for $L=1600$ respectively.
The proposed changes are biased towards $\Delta x>0$,
because here, in the negative tail, a Monte-Carlo step $\Delta x>0$
corresponds to a move where the probability $P(x)$ is bigger.
The histogram of the accepted moves is on the contrary
almost symmetric around $\Delta x =0$ in order to generate
a non-biased random walk. For $\Delta x <0$, the two histograms
almost coincide, i.e. a move $\Delta x<0$ is almost always accepted.
As the size $L$ grows, the proposed moves in the relative variable
$x$ are smaller, and as a consequence, 
the acceptation rate grows with $L$.

\begin{figure}[htbp]
\includegraphics[height=6cm]{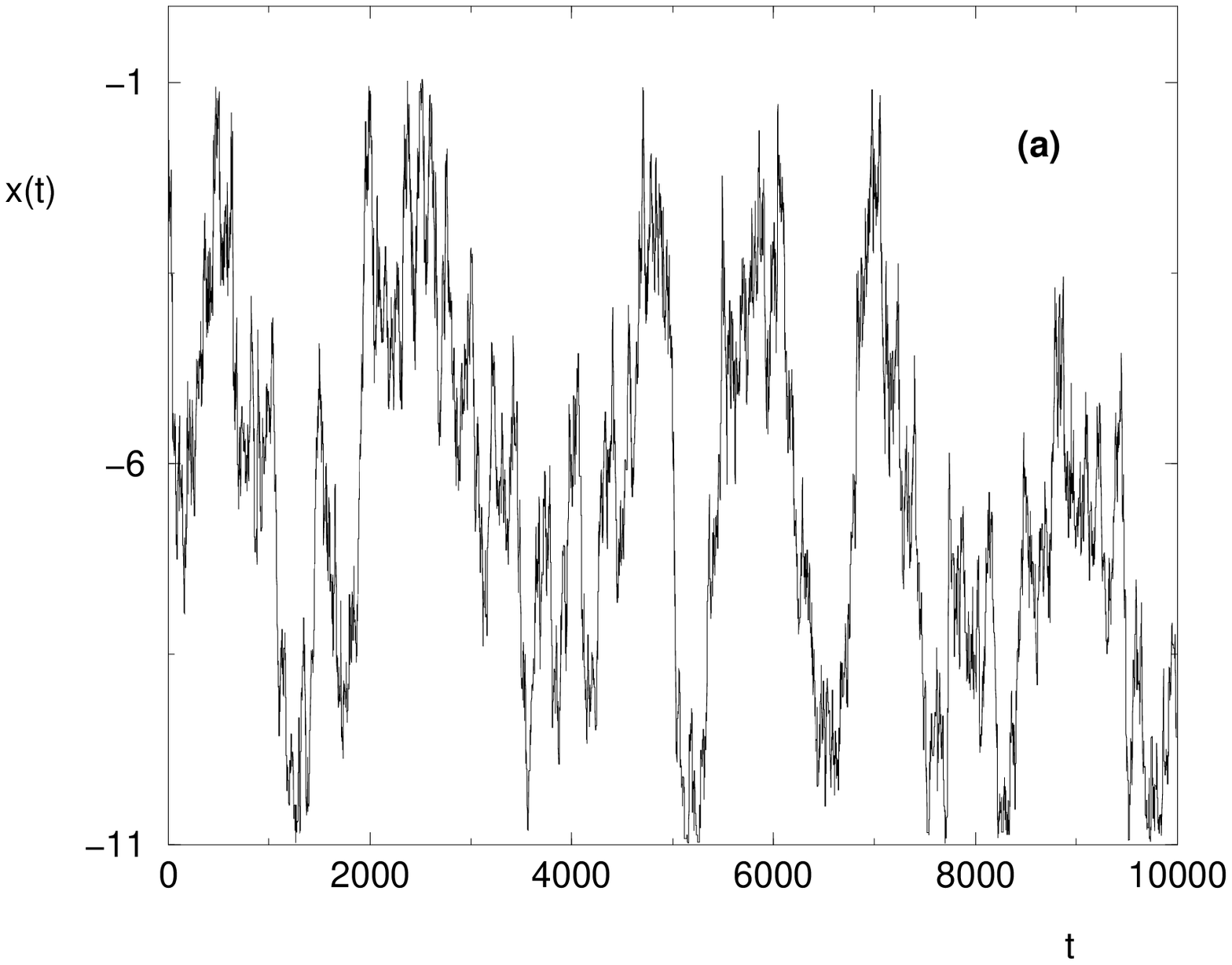}
\hspace{1cm}
\includegraphics[height=6cm]{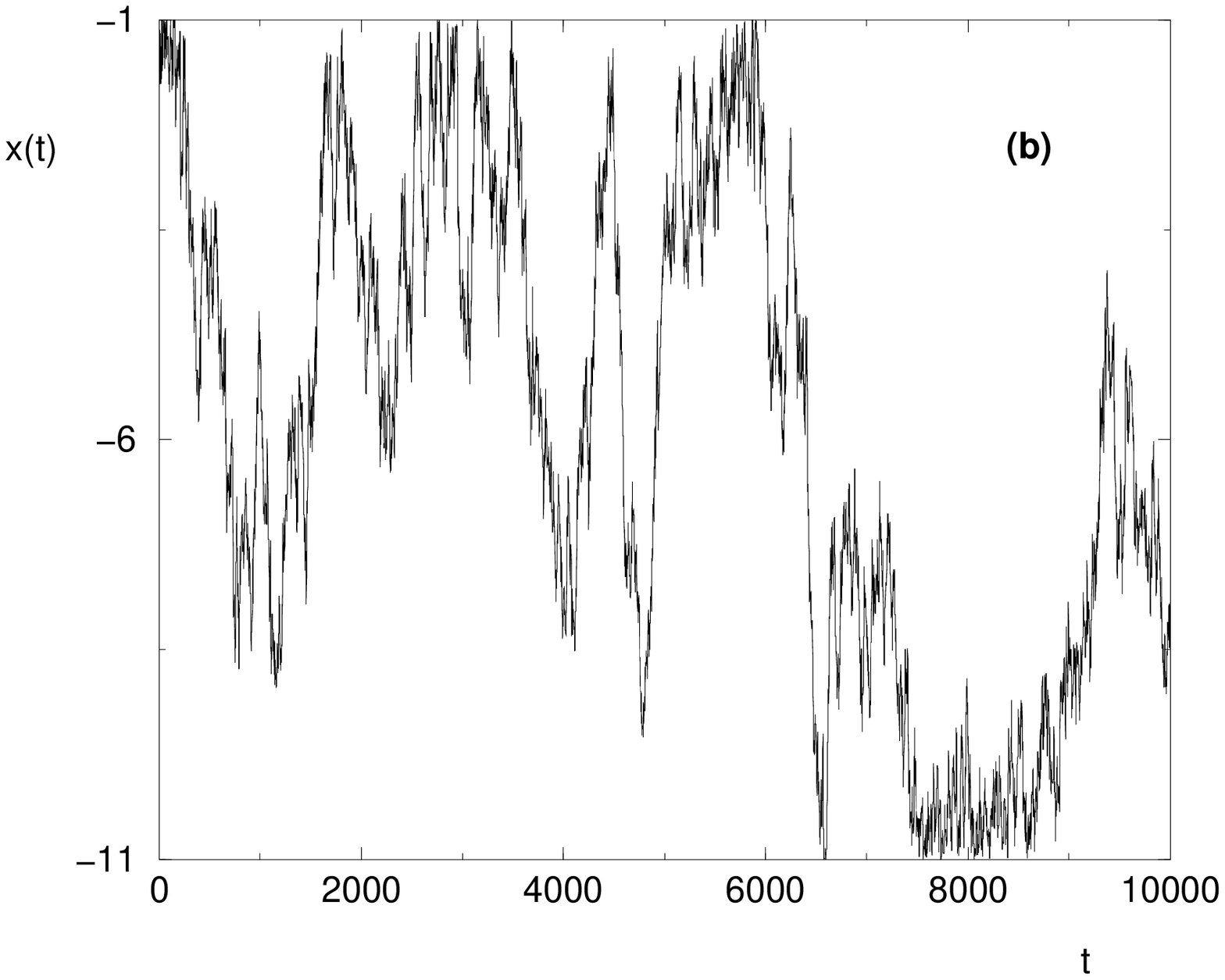}
\caption{ (Color online)  Monte-Carlo procedure to measure
 the negative tail on $x \in [-11,-1]$ in $d=1$ :
process $x(t)$ during the first $1 \leq t \leq 10 000$ Monte-Carlo iterations 
(a) for $L=50$ 
(b)  for $L=200$  }
\label{fi1d15}
\end{figure}

The resulting process $x(t)$ are shown on Fig. \ref{fi1d15}
for the first $1 \leq t \leq 10 000$ Monte Carlo iterations,
for $L=50$ and $L=200$ respectively.
The time $t_{cross}$
needed to cross the interval $[x_{min},x_{max}]=[-11,-1]$
grows with $L$.

\subsection{ Convergence towards the exact Tracy-Widom distribution }

\begin{figure}[htbp]
\includegraphics[height=6cm]{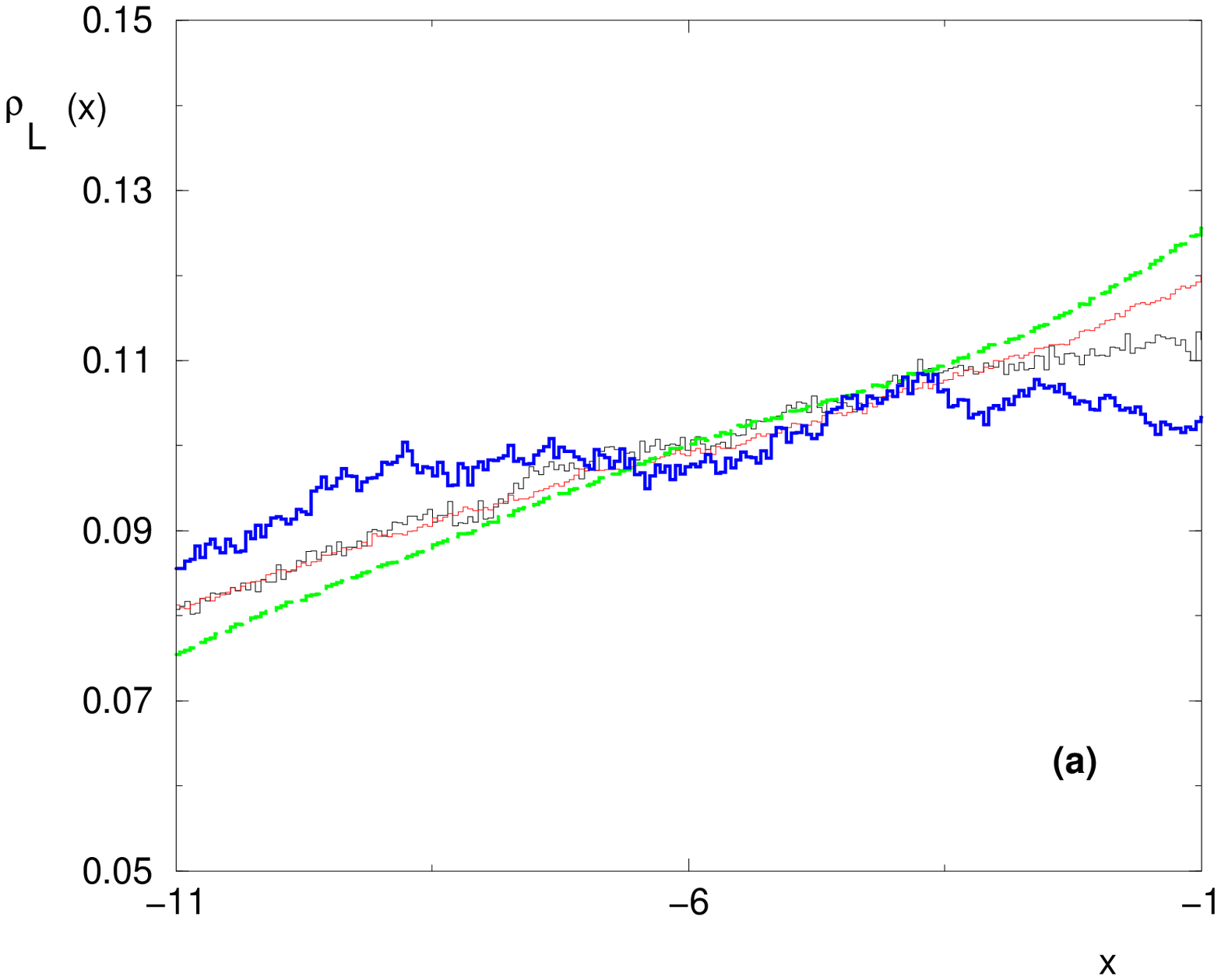}
\hspace{1cm}
\includegraphics[height=6cm]{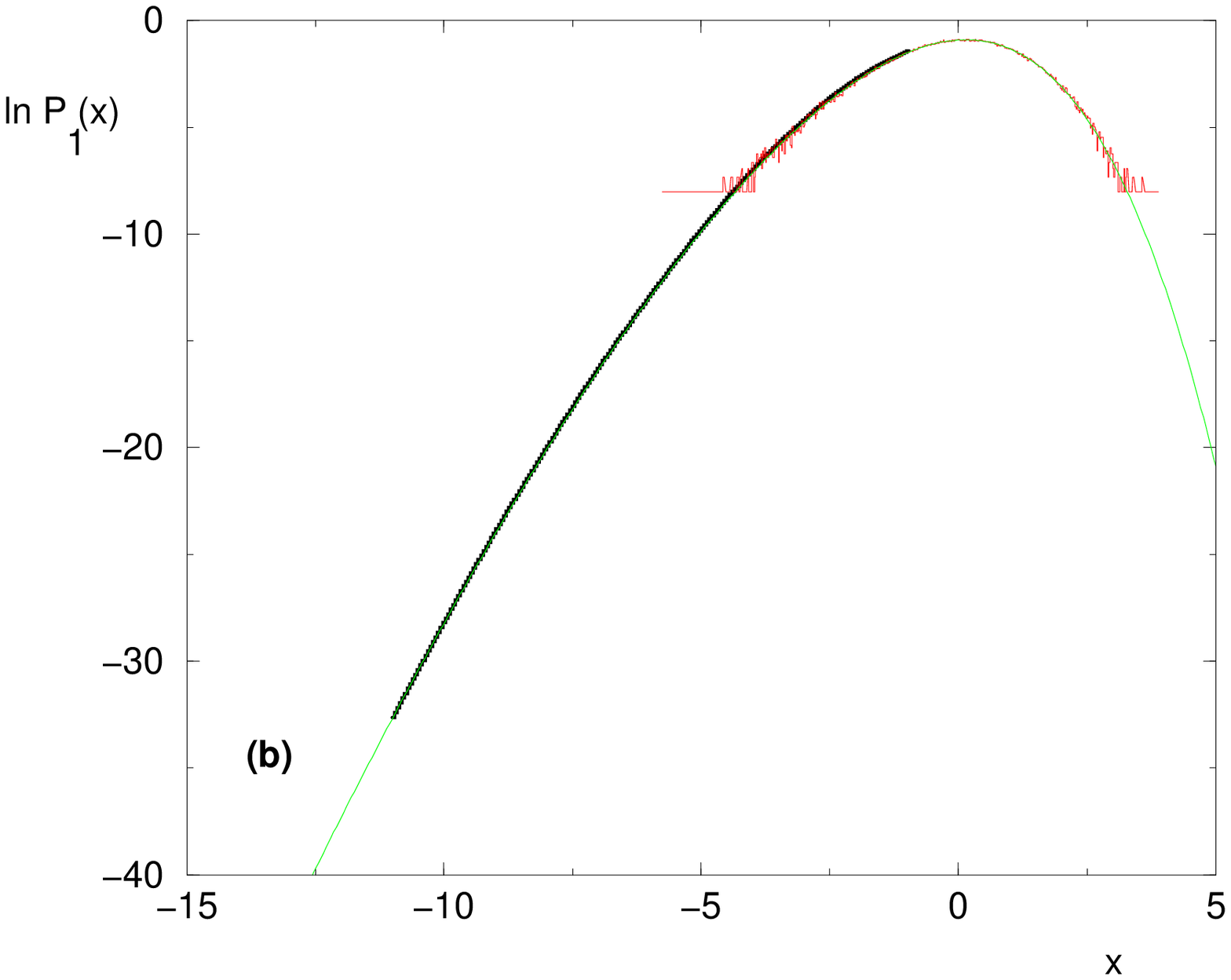}
\caption{ (Color online)  Monte-Carlo procedure to measure
 the negative tail on $x \in [-11,-1]$ in $d=1$ :
(a) relative histogram $\rho_L(x)=P_L(x)/P_{TW}^{GOE}(x)$ with
respect to the exact guiding function :
convergence towards the flat histogram as $L$ grows : $L=200$ (dashed line), $400, 800, 1600$ (thick line). 
(b) logarithmic plot of the negative tail of the probability distribution
$P_1(x)$, as compared to simple sampling result
 for $L=1600$. The exact Tracy-Widom distribution is also shown (thin line)
to demonstrate the validity of the Monte-Carlo procedure.   }
\label{fi1d3545}
\end{figure}

On Fig. \ref{fi1d3545} (a), we show the relative histogram
$P_L(x)/P_{TW}^{GOE}(x)$ of the measured $P_L(x)$
via the Monte Carlo procedure as $L$ grows
with respect to the Tracy-Widom GOE distribution that represents the
asymptotic exact result for $L \to \infty$ : these relative histograms
becomes flatter as $L$ grows. 

On Fig. \ref{fi1d3545} (b), we show for comparison : \\
(i) the simple sampling histogram for $L=1600$ \\
(ii) the importance sampling measure of the tail on $x \in [-11,-1]$ 
 for $L=1600$ \\
(iii) the exact Tracy-Widom GOE distribution.

Our conclusion is thus that the Monte-Carlo in the disorder
is a very efficient method to probe accurately the tails,
since they allow to reproduce the exact result on the range
$x \in [-11,-1]$ for sizes up to $L=1600$.

\subsection{ Extraction of the negative tail exponent value }

\label{charbon}

 Let us now make some comments on the extraction
 of the negative tail exponent value.
 The exact
Tracy-Widom GOE distribution $P_{TW}^{GOE}(x)$
has for negative exponent $\eta_1=3/2$.
However, the following fits of this distribution $P_{TW}^{GOE}(x)$
on the finite range $x \in [-11,-1]$ give slightly larger values : \\
(i) the fit of $\left( \ln P_{TW}^{GOE}(x) \right) $ by $ a - b (-x)^{\eta_1}$ 
containing three parameters yields $\eta_1 \sim 1.58$ \\
(ii)  the fit of $\left( \ln P_{TW}^{GOE}(x)\right) $
 by $ a - b (-x)^{\eta_1}+c \ln (-x)$ 
containing four parameters yields $\eta_1 \sim 1.54$. \\
This shows that the extracted value of the negative tail exponent
from data on the finite range $x \in [-11,-1]$ is not very precise
if there is not information on the subleading terms.

Similarly in higher $d$ below, we expect that 
the Monte-Carlo procedure gives very accurate data on the 
range where the tail is measured, but that the extraction of the 
negative tail exponent value suffers from some error directly related
to the range that is probed.

\section{ Results for the ground state energy distribution in $d=2$}

\label{res2d}

\subsection{ Numerical details }

In dimension $d=2$, we have chosen to work on the interval
$ [x_{min},x_{max}]=[-10.,-1.]$ for the rescaled variable $x$ 
(Eq. \ref{scalinge0}), i.e. to probe the negative tail up to probabilities
of order $P_1(x) > 10^{-23}$. 
We now give the sizes $L$
we have studied, together with the standart deviation
 $\Delta E_0(L)$
measured by simple sampling and used in the rescaling of Eq. 
\ref{scalinge0} (the averaged values $E_0^{av}(L)$ can be found in 
our previous work \cite{DPexcita}),
the corresponding 
number $T_L$ of Monte-Carlo iterations,
the acceptation rate 
$\tau_{acc}(L)$ of Monte-Carlo moves
\begin{eqnarray}
L && =20,40,80, 120,160 \\
\Delta E_0(L) && \sim 1.58, 1.85,2.18,2.40, 2.54 \\
T_L && = 125.10^7, 27.10^7, 47.10^6, 34.10^5, 64.10^4 , 21.10^4 \\
\tau_{acc}(L) && \sim  0.24, 0.28, 0.38,0.47, 0.5, 0.54
 \end{eqnarray}

\subsection{ Monte-Carlo results }

\begin{figure}[htbp]
\includegraphics[height=6cm]{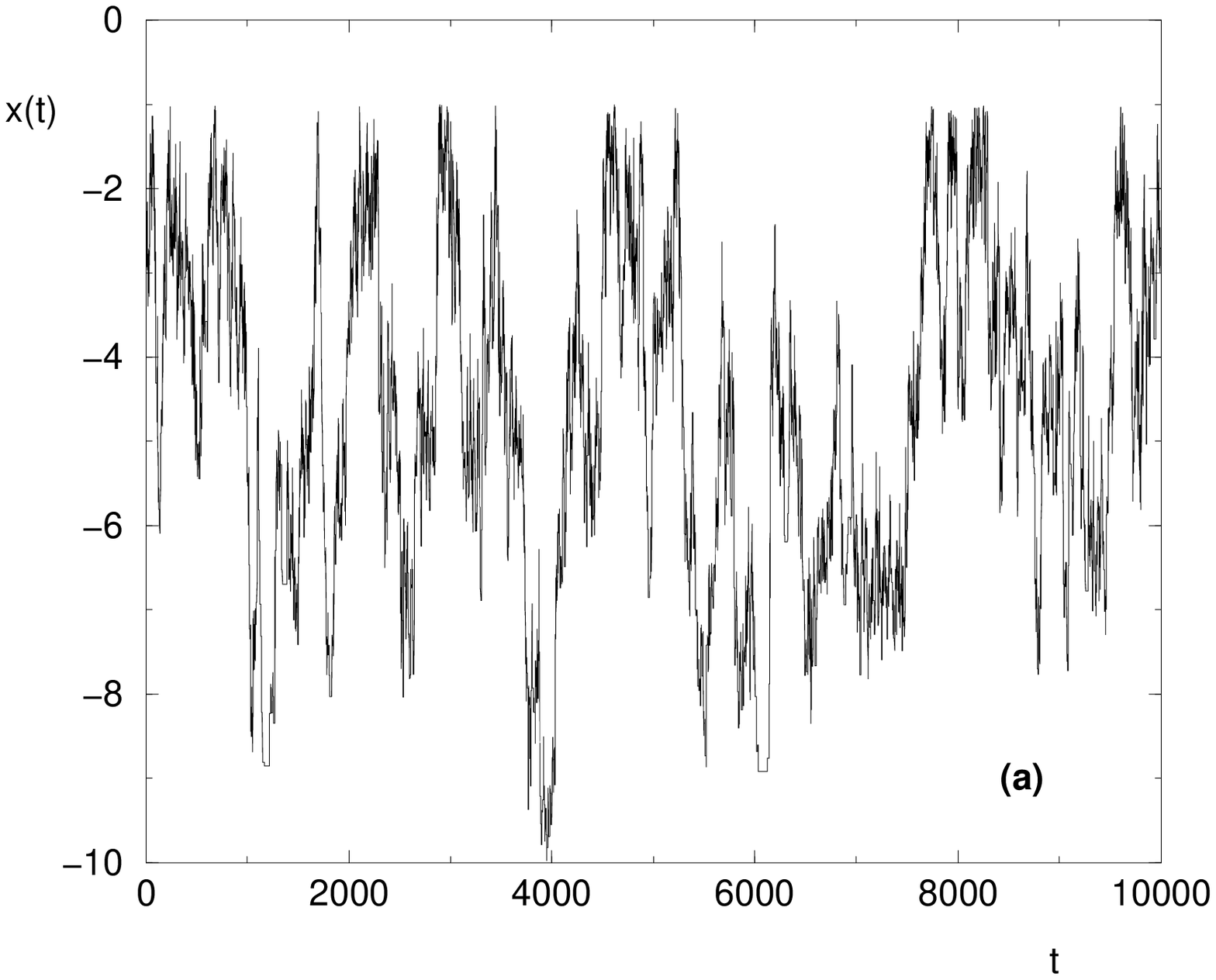}
\hspace{1cm}
\includegraphics[height=6cm]{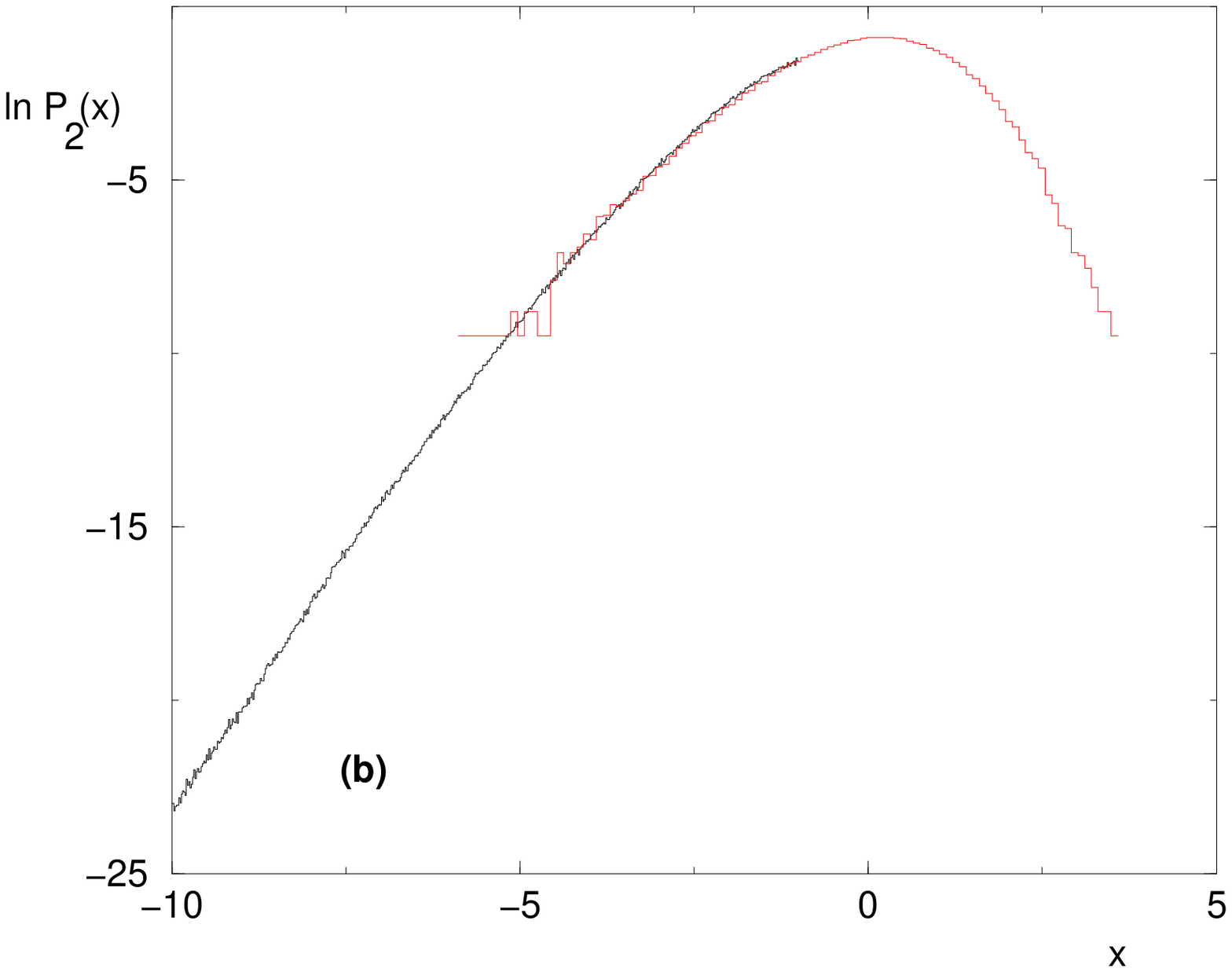}
\caption{ (Color online)  Monte-Carlo procedure to measure
 the negative tail on $x \in [-10,-1]$ in $d=2$ for $L=120$ :
(a) process $x(t)$ during the first $1 \leq t \leq 10 000$
 Monte-Carlo iterations
(b) logarithmic plot of the negative tail of the probability distribution
$P_2(x)$, as compared to simple sampling result. }
\label{fig2d}
\end{figure}

On Fig. \ref{fig2d} (a), we show the process $x(t)$
during the first 10 000 Monte-Carlo iterations for $L=120$.
On Fig. \ref{fig2d} (b), we compare the importance sampling measure
of the negative tail with respect to the simple sampling evaluation. 

\subsection{ Negative tail exponent $\eta_{d=2}$ }

From the point of view of the convergence in $L$ towards a fixed
distribution, we find that the negative tail measured for the two bigger
sizes $L=120$ and $L=160$ nearly coincide on the whole interval
 $ [x_{min},x_{max}]=[-10.,-1.]$ under study
( whereas our results for the smaller sizes do not).

As explained previously in Section \ref{charbon}
for the case $d=1$, the error on the estimated value
of the negative tail exponent is due to the 
range $ [x_{min},x_{max}]=[-10.,-1.]$ over which the fits are made.
As in Section \ref{charbon}, we have tried to fit our result for
$\left( \ln P_2(x) \right)$ as measured for the sizes $L=120$ and $L=160$ by
the two following fits, with or without power-law corrections
with respect to the leading exponential term :
 \\
(i) the first fit $ a - b (-x)^{\eta_2}$ 
containing three parameters yields $\eta_2 \sim 1.4$ \\
(ii)  the second fit  by $ a - b (-x)^{\eta_1}+c \ln (-x)$ 
containing four parameters yields $\eta_2 \sim 1.3$. \\

Our conclusion is thus that the extracted value of the negative tail exponent
from our data on the finite range $x \in [-10,-1]$ is not very precise
in the absence of information on the subleading terms,
but is compatible with the value $\eta_2^Z=1.32$ predicted
by Zhang's argument (see Eqs. \ref{etad} and \ref{etad2et3}).

\section{ Results for the ground state energy distribution in $d=3$}

\label{res3d}

\subsection{ Numerical details }

In dimension $d=3$, we have chosen to work on the interval
$ [x_{min},x_{max}]=[-10.,-1.]$ for the rescaled variable $x$ 
(Eq. \ref{scalinge0}), i.e. to probe the negative tail up to probabilities
of order $P_1(x) > 10^{-21}$. 
We now give the sizes $L$
we have studied, together with the standart deviation
 $\Delta E_0(L)$
measured by simple sampling and used in the rescaling of Eq. 
\ref{scalinge0} (the averaged values $E_0^{av}(L)$ can be found in 
our previous work \cite{DPexcita}), the corresponding 
number $T_L$ of Monte-Carlo iterations,
the acceptation rate 
$\tau_{acc}(L)$ of Monte-Carlo moves,
and the number $n_{cross}(L)$ of crossings of the box $ [x_{min},x_{max}]=
[-10.,-1.]$.
\begin{eqnarray}
L && = 12,24,36,48,60,72\\
\Delta E_0(L) && \sim 1.15, 1.30,1.39,1.46, 1.52 ,1.55 \\
T_L && = 64.10^6, 43.10^5, 75.10^4, 95. 10^4, 34.10^4, 182.10^3  \\
\tau_{acc}(L) && \sim 0.24, 0.27,0.32, 0.34, 0.36, 0.37 \\
n_{cross} (L) && \sim  20 000 , 2400,300, 522, 166,  94
 \end{eqnarray}

\subsection{ Monte-Carlo results }

\begin{figure}[htbp]
\includegraphics[height=6cm]{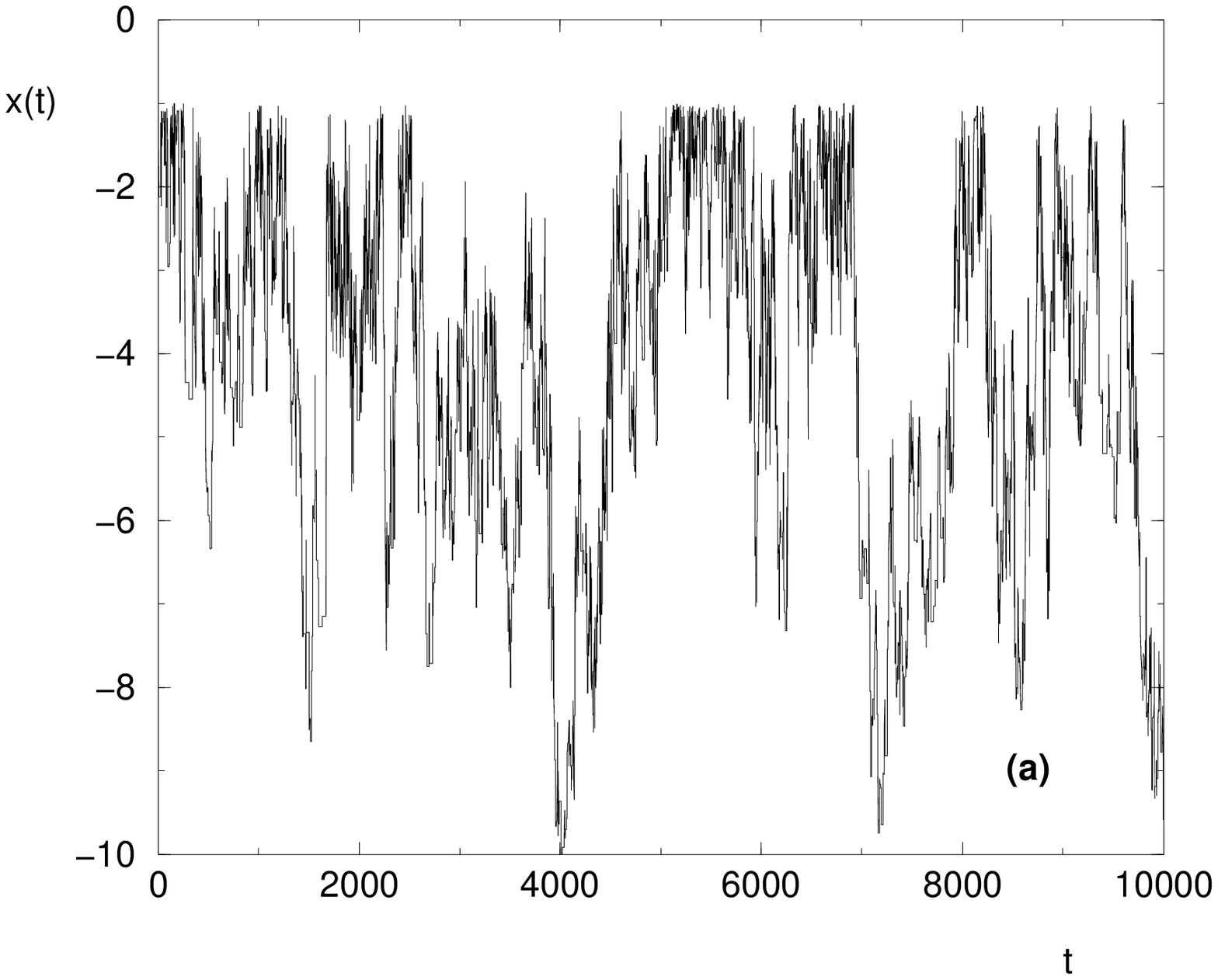}
\hspace{1cm}
\includegraphics[height=6cm]{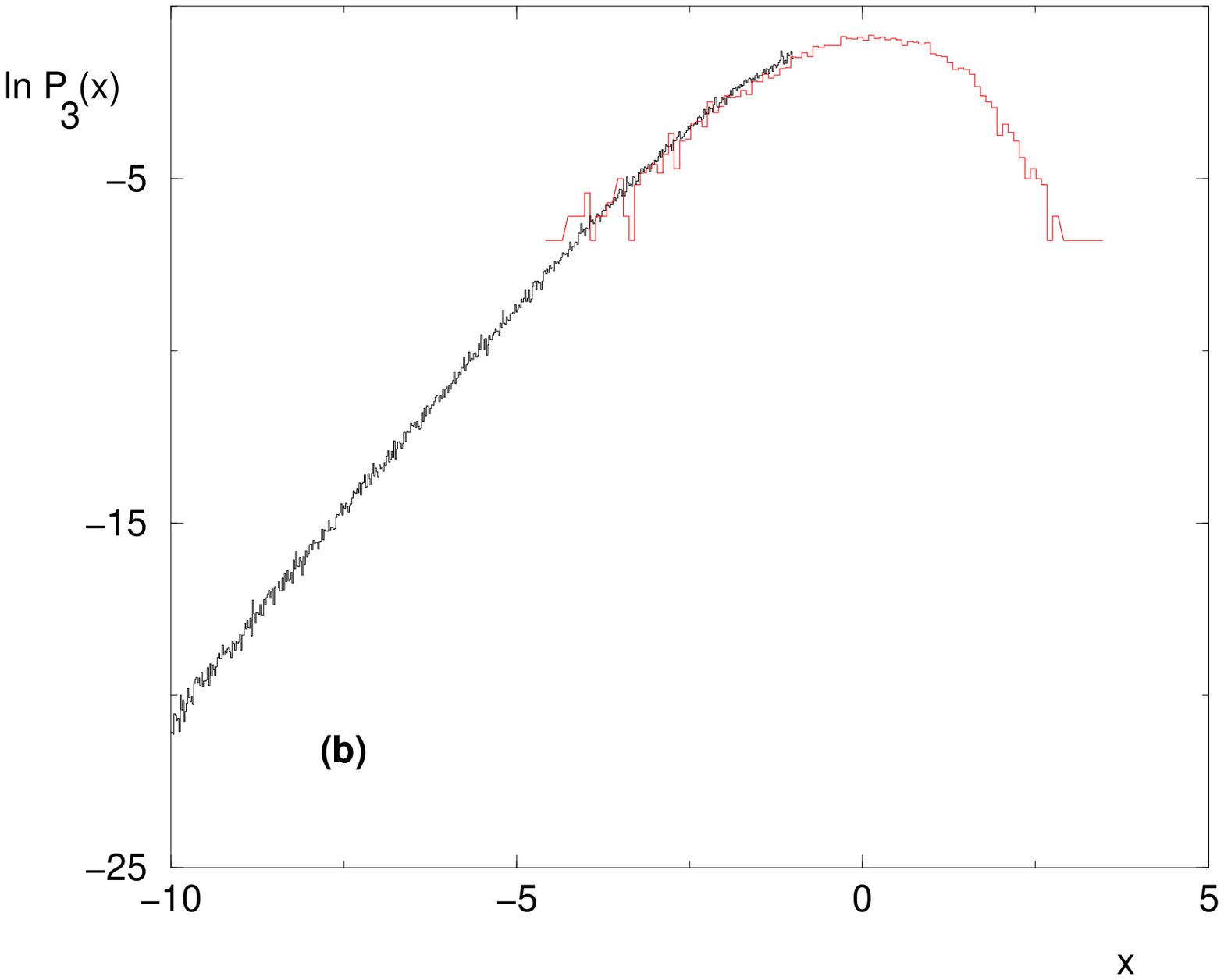}
\caption{ (Color online)  Monte-Carlo procedure to measure
 the negative tail on $x \in [-10,-1]$ in $d=3$ for $L=72$ :
(a) process $x(t)$ during the first $1 \leq t \leq 10 000$
 Monte-Carlo iterations
(b) logarithmic plot of the negative tail of the probability distribution
$P_3(x)$, as compared to simple sampling result. }
\label{fig3d}
\end{figure}

On Fig. \ref{fig3d} (a), we show the process $x(t)$
during the first 10 000 Monte-Carlo iterations for $L=72$.
On Fig. \ref{fig3d} (b), we compare the importance sampling measure
of the negative tail with respect to the simple sampling evaluation.

\subsection{ Negative tail exponent $\eta_{d=3}$  }

As in Section \ref{charbon}, we have tried to fit our result for
$\left( \ln P_3(x) \right)$  by
the two following fits, with or without power-law corrections
with respect to the leading exponential term :
 \\
(i) the first fit $ a - b (-x)^{\eta_3}$ 
containing three parameters yields $\eta_3 \sim 1.25 $ \\
(ii)  the second fit  by $ a - b (-x)^{\eta_1}+c \ln (-x)$ 
containing four parameters yields $\eta_3 \sim 1.15$. \\

Our conclusion is that the extracted value of the negative tail exponent
from our data on the finite range $x \in [-10,-1]$ is not very precise
in the absence of information on the subleading terms,
but is compatible with the value $\eta_3^Z=1.23$ predicted
by Zhang's argument (see Eqs. \ref{etad} and \ref{etad2et3}).

\section{ Conclusion }

\label{conclusion}

In this paper, we have adapted the
importance-sampling method in the disorder
proposed in \cite{Ko_Ka_Ha} for spin-glasses,
to measure with high precision the negative tail of the 
ground-state energy distribution $P_d(E_0)$
for the directed polymer in a random medium of dimension $d=1,2,3$.
In $d=1$, we have checked the validity 
of the procedure by a direct comparison
with the exact result, namely the Tracy-Widom GOE distribution.
In dimensions $d=2$ and $d=3$, we have measured the negative tail up to
$P \sim 10^{-22}$.
Our results are in agreement with Zhang's argument,
stating that the negative tail exponent $\eta(d)$ of
the asymptotic behavior
 $\ln P (E_0) \sim - \vert E_0 \vert^{\eta(d)}$ as $E_0 \to -\infty$
is directly related to the fluctuation exponent $\theta(d)$
via the simple formula $\eta(d)=1/(1-\theta(d))$.

Along the paper, we have also
discussed the similarities and differences with
spin-glasses. In particular, we have argued that 
the application of Zhang's argument for 
the Sherrington-Kirpatrick model of spin-glasses
points towards an asymptotic distribution which is 
not a generalized Gumbel
distribution $g_m(x)$, in contrast with the current way of fitting
the numerical data \cite{palassini,Ko_Ka_Ha}, but involves instead
some non-trivial negative tail exponent $\eta_{SK}>1$
directly related to the fluctuation exponent (Eq. \ref{etaSK}).
The fact that the fitting value $m$ of
generalized Gumbel
distribution $g_m(x)$ depends
on the probed range in the variable $x$ ($m \sim 6$ via simple sampling
\cite{palassini}  and $m \sim 11$ via importance sampling
\cite{Ko_Ka_Ha}) also points towards $\eta_{SK} >1$.
More generally, we have explained in details how fits with generalized
Gumbel distributions of the core of the distribution
could be very misleading if one is interested on the tails,
since all Gumbel distributions correspond to the exponent $\eta=1$,
which is very restrictive.

Finally, our conclusion concerning the algorithm is that the 
importance sampling
Monte-Carlo Markov chain in the disorder introduced in \cite{Ko_Ka_Ha} 
is a very efficient method to probe precisely
the tails of probability distributions over the samples.
In the field of disordered systems, this Monte Carlo procedure will be
very useful to study probability distributions of other observables,
beside the ground state energy.

\end{document}